\documentclass[10pt]{IEEEtran}
\usepackage{amsmath}
\usepackage[T1]{fontenc}
\usepackage{algorithm}
\usepackage{alphalph}
\usepackage{algorithmic}
\usepackage{listings}
\usepackage{ulem} \usepackage{cancel}
\usepackage{nomencl}
\makenomenclature

\usepackage{amsmath}
\usepackage{graphicx}
\usepackage{color}
\usepackage[cmintegrals]{newtxmath}
\usepackage[caption=false,font=footnotesize]{subfig}
\ifCLASSOPTIONcompsoc
\usepackage[caption=false,font=normalsize,labelfont=sf,textfont=sf]{subfig}
\else
\usepackage[caption=false,font=footnotesize]{subfig}
\fi
\usepackage[colorlinks,citecolor=red,urlcolor=blue,bookmarks=false,hypertexnames=true]{hyperref} 
\usepackage{caption}
\usepackage{lipsum}
\usepackage{nameref}
\usepackage{varioref}
\usepackage{hyperref}
\usepackage{setspace}

\newtheorem{proposition}{Proposition}
\newtheorem{corollary}{Corollary}
\begin{document}
\title{Adversarial Robustness in Cognitive Radio Networks}
\author{Makan~Zamanipour,~\IEEEmembership{Member,~IEEE}
\thanks{Makan Zamanipour is with Lahijan University, Shaghayegh Street, Po. Box 1616, Lahijan, 44131, Iran, makan.zamanipour.2015@ieee.org. Copyright (c) 2015 IEEE. Personal use of this material is permitted. However, permission to use this material for any other purposes must be obtained from the IEEE by sending a request to pubs-permissions@ieee.org. }
}

\maketitle
\markboth{IEEE, VOL. XX, NO. XX, X 2021}%
{Shell \MakeLowercase{\textit{et al.}}: Bare Demo of IEEEtran.cls for Computer Society Journals}
\begin{abstract}
\textit{When an adversary gets access to the data sample in the adversarial robustness models and can make data-dependent changes, how has the decision maker consequently, relying deeply upon the adversarially-modified data, to make statistical inference? How can the resilience and elasticity of the network be literally justified $-$ if there exists a tool to measure the aforementioned elasticity?} The principle of byzantine resilience distributed hypothesis testing (BRDHT) is considered in this paper for  cognitive radio networks (CRNs) $-$ without-loss-of-generality, something that can be extended to any type of homogeneous or heterogeneous networks $-$ while the byzantine primary user (PU) has a signal-to-noise-ratio (SNR) including the coefficient of $\frac{d\ell  \big ( \theta | \mathscr{s}_0 \big )}{d\ell  \big ( \theta  \big )} $ which is in relation to the temporal rate of the $\alpha-$leakage as the appropriate tool to measure the aforementioned resilience. Our novel online algorithm $-$ which is named $\mathbb{OBRDHT}$ $-$ and solution are both unique and generic over which an evaluation is finally performed by simulations $-$ e.g. an evaluation of the total error as the false alarm probability in addition to the miss detection probability versus the sensing time.
\end{abstract}


\maketitle

\IEEEdisplaynontitleabstractindextext
\IEEEpeerreviewmaketitle


\section{Introduction} 

\IEEEPARstart{A}ccurate and efficient procedures are required to find some unique ans specific information sets e.g. distributions of data sets or the number of libraries in the controlled network. However, the controlled network may be of a technically adversarial situation in which the users are able to be potentially subjected to some knock-on effects $-$ resulting in an a priori unknown subset of Byzantines\footnote{Those users who do data falsification $-$ something that happens in the physical-layer.} who purposefully follow totally different trajectories. 

Aim at classifying the Byzantines, each good user should be collaboratively required to repeatedly share their beliefs, i.e., observations with others $-$ since in reality, only a single good agent cannot be useful for the aforementioned classification, due to practical limitations such as a noisy plant. However, there arises an important question that how much reliable the aforementioned collaborative sharing is, as Byzantines undoubtedly have a potential to arbitrarily share over-written information aimed at preventing of being infered. 

Since every possible subset relating to Byzantines
is considered a hypothesis, the classification issue expressed above is called distributed
hypothesis testing $-$ where a subset of users cooperatively identifies the unknown true
hypothesis according to their local beliefs.

Distributed networks have mainly two valid beneficiaries compared with centralised systems while the former ones: (\textit{i}) can technically guarantee a resilience to fraud; and (\textit{ii}) can tolerate failures. Meanwhile, distributed systems have three main features: (\textit{i}) Agreement, that is, an overall decision on the same value by all true processes; (\textit{ii}) termination, that is, to make a decision in an expected finite time-zone by all true processes; and (\textit{iii}) validity, that is, the agents who are in consensus never leave their agreement. 

Inversely, aimed at preventing the consensus to occur, all Byzantine processes are controlled by an Adversary. In fact, while binary byzantine consensus degrades the overall performance, distributed solutions where each agent communicates to its neighbors without a fusion center can enrich the resilience.

Distributed algorithms rely mostly upon \textit{beliefs} which are probability distributions over the hypotheses. Now, each agent’s actual belief should theoretically converge to the true hypothesis in order to guarantee the resilient against the bad agents who arbitrarily share modified information.

\begin{figure}[t]
\centering
\subfloat{\includegraphics[trim={{15mm} {32mm} {228mm} {4mm}},clip,scale=0.46]{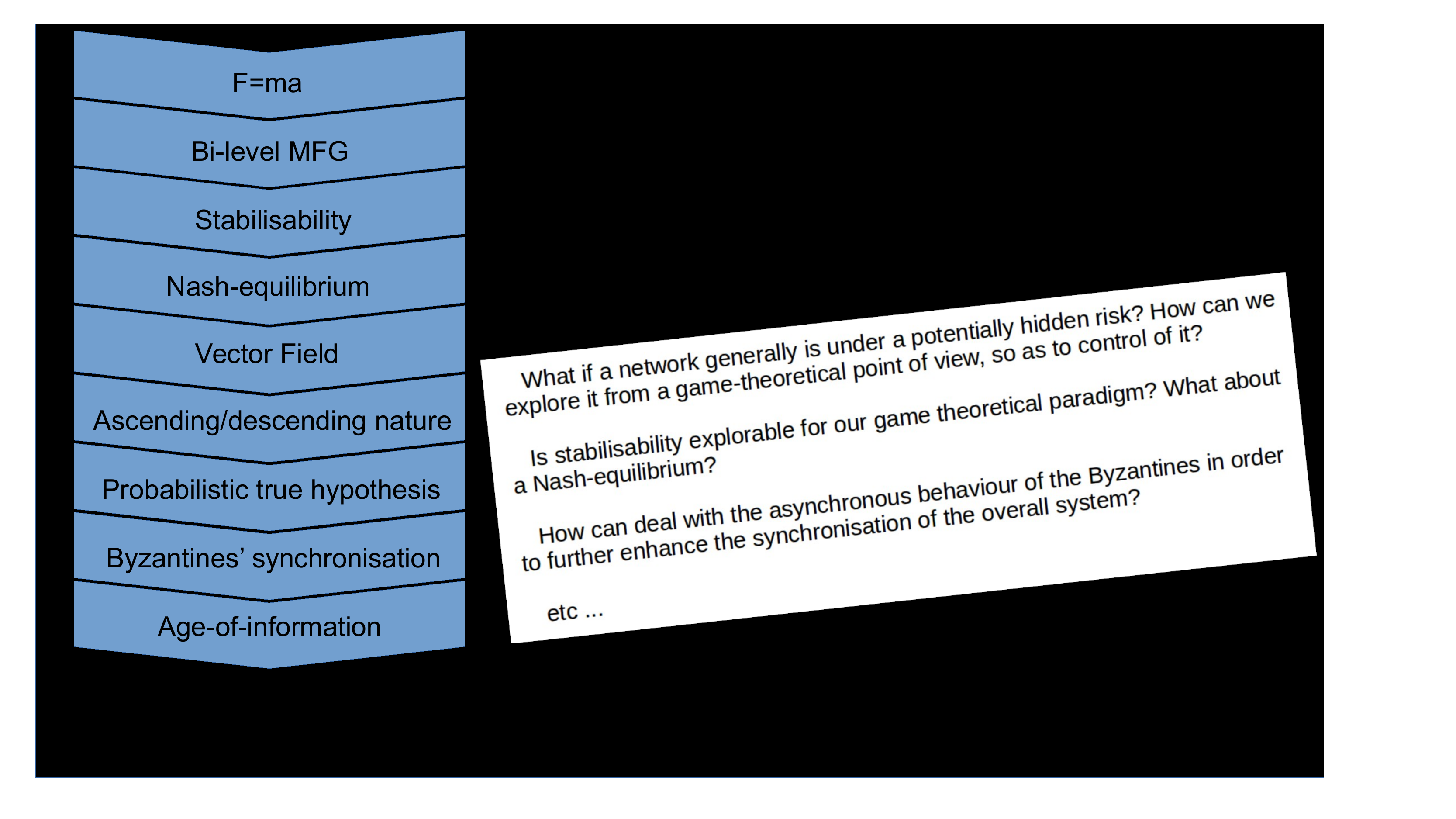}} 
\caption{Flow of the problem-and-solution and our contribution in this paper. } \label{F0}
\label{fig:EcUND} 
\end{figure}

\subsection{Related work}

In this part, we explore the related work recently appeared in the literature. 

\subsubsection{Byzantine resilience distributed hypothesis testing (BRDHT) \cite{momi1}-\cite{8}} 

In \cite{momi1, momi2, momi3, momi4, momi5, momi6, momi7, momi8, momi9, momi10}, some novelties from either practical or theoretical or both points of view were proposed for real world applications such as: the Internet of Things, the multi-sinks, the one-bit double quantisation algorithms, multi-cast routing in multi-hop wireless networks, decentralized learning, vehicular social networks or even cognitive radio networks. In \cite{zz1}, an unknown subset of auctioned agents was considered that may consciously share modified information to diminish the global optimisation. While each agent try to maintain and update two sets of beliefs, namely actual and local and actual ones, two distributed algorithms were consequently proposed in \cite{zz1}. A sparse encoding framework was proposed in \cite{1}, while a trade-off was defined between the resource requirements and the \textit{corruption threshold.} in \cite{2}, the jamming behaviour of Byzantines were considered for the first time. The work in \cite{3} as the first research to byzantine-resilient optimisation in which no leader player existed for coordination, the authors found the global optima in the context of a convex fashion. In \cite{4} as the first single-server byzantine-resilient scheme for secure federated learning based on an integrated stochastic quantisation, some important trade-offs were examined for the network size, privacy protection and user dropouts. In \cite{5}, the asymptotically learning issue of the true hypothesis in an almost sure sense was explored. In \cite{6}, it was perfectly proven that the probability of sample paths through which the rejection rate of false observations tangentially passes from the mean-rate, would theoretically experience an exponential plummet. Under the assumptions that: (\textit{i}) the number of Byzantines is locally upper-bounded, and (\textit{ii}) each good agent neglects the most extreme values on his local vicinity consequently altering his observations, some novel and secure ideas were well-developed in \cite{7} for both types of first-order and second-order systems. Instead of the case that agents update their observations according to a min-rule, a novel distributed learning rule was proposed in \cite{8}, which does not concentrate on “belief-averaging”. In fact, the research in \cite{8} works in the sense that each agent is able to learn the true hypothesis asymptotically almost surely. It was also proven that each falsification is ruled out by every agent exponentially fast: (\textit{i}) with probability 1, and (\textit{ii}) at an enhanced network-independent rate.

\subsubsection{Byzantine attacks in CRNs \cite{bim1}-\cite{bim5}}
In \cite{bim1}, the issue of optimal sensing disruption for a cognitive radio adversary was technically explored. In \cite{bim2}, a secure and energy-efficient cooperative spectrum sensing scheme was newly proposed. In \cite{bim3}, a generalized framework for a power-limited adversary in terms of the optimal sensing disruption was theoretically proposed. In \cite{bim4} and for the process of spectrum sensing, secure routing scenarios were deeply taken into account which was based mainly upon the nodes' behaviour. In \cite{bim5} and with the aid of the Thompson sampling technique, an online primary user emulation attack was investigated. 

\subsubsection{Asynchronous Byzantine consensus \cite{s}-\cite{a4}} 

The issue of asynchronous byzantine agreement was taken theoretically into account for incomplete networks in \cite{s}. In \cite{ss}, through a weak round installer, a highly effective binary byzantine faulty tolerant consensus algorithm was proposed. In the aforementioned algorithm proposed in \cite{ss}, an initial broadcast was fulfilled by non-faulty agents. In \cite{a1}, through asynchronous message-passing networks which may include the issue of byzantine failure, distributed causal shared memory in the client-server model was considered. The issue of lattice agreement was investigated in asynchronous systems in \cite{a2} where each process would make a decision over an metric from a per-defined lattice under the assumption of the problem of byzantine failure. In \cite{a3}, the problem of online payments was explored, where resources were transferable among agents. Towards such end, \textit{Astro}, a system was highly effectively proposed to make a solution in a completely asynchronous manner. It was also proven that Astro can enable an adequately more effective implementation than consensus-based frameworks. In \cite{a4}, it was proposed a novel model by which one can use those extra founds being able to be created in byzantine fault-tolerant protocols in asynchronous networks. The aforementioned model aimed at partitioning the nodes into clusters and concurrently and independently processing transactions on different partitions. 

\subsubsection{Information theory \& game-theory \cite{9}-\cite{11}} 

From an information theoretic game-theoretical point of view, although some rare work has been done e.g. \cite{9, 10, 11}, this dominant area of research is still open.
In \cite{9}, and for general alphabets, the saddle-point of the conditional Kullback-Leibler-Divergence to the conditional Re'nyi divergence was strongly evaluated. In \cite{10}, the information pattern of a mean-field-theoretical scheme was evaluated. In \cite{11}, a trade-off was physically realised in the context of ''how much fast and how much secure'' w.r.t. two essential limits in mean-field-games (MFGs). 

\begin{figure}[t]
\centering
\subfloat{\includegraphics[trim={{61mm} {40 mm} {86mm} {50mm}},clip,scale=0.4]{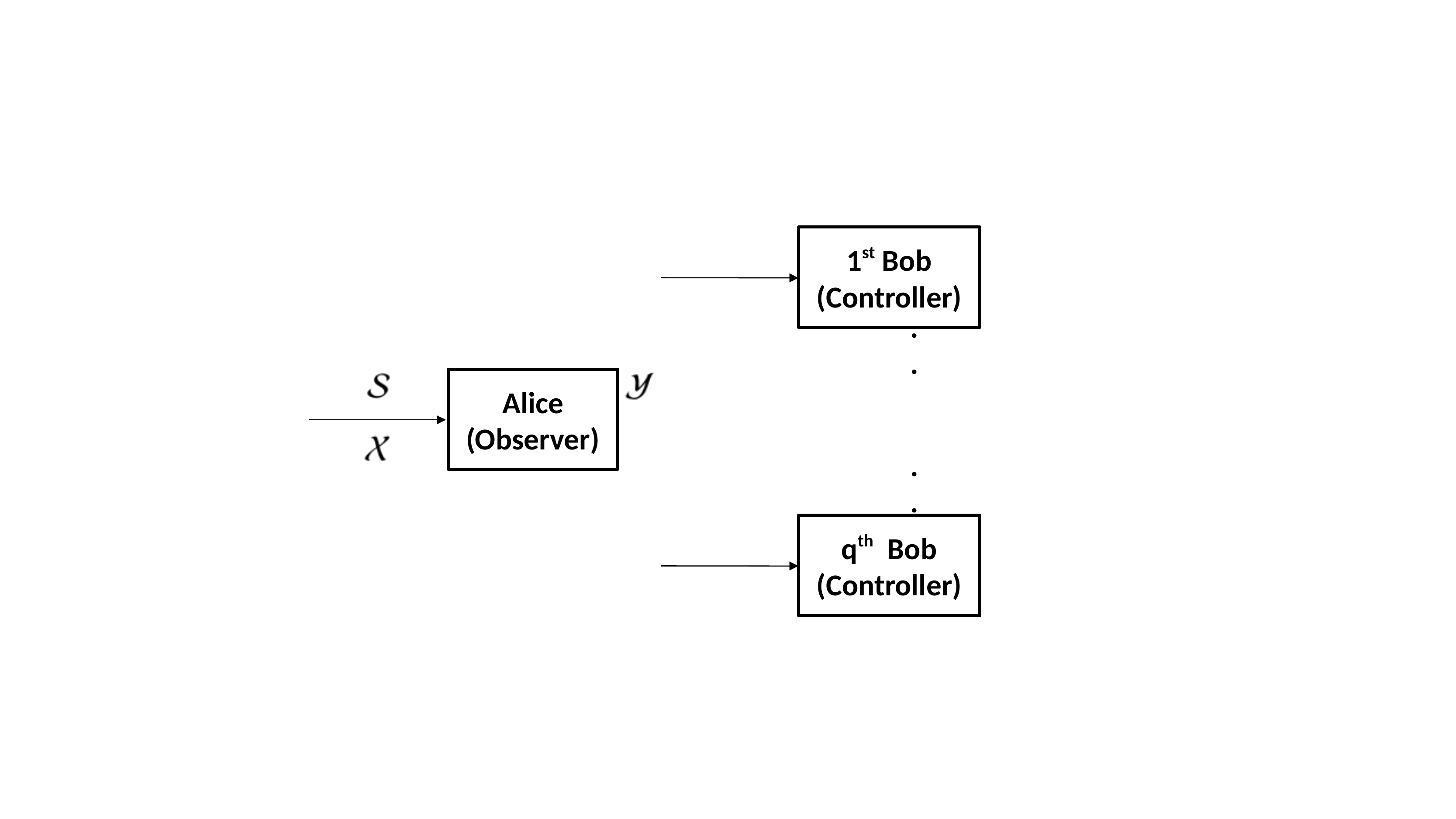}} 
\caption{Distributed hypothesis testing model.} \label{dool}
\label{fig:EcUND} 
\end{figure}

\subsection{Motivations and contributions}
In this paper, we are interested in responding to the following questions: \textit{What if a network generally is under a potentially hidden risk? How can we explore it from a game-theoretical point of view, so as to control of it? Is stabilisability explorable for our game theoretical paradigm? How can deal with the asynchronous behaviour of the Byzantines in order to further enhance the synchronisation of the overall system?} With regard to the incomplete version of the literature, the questions expressed here strongly motivate us to find an interesting solution, according to which our contributions are described in the following.

While the data set is distributively compressed and transfered to an estimator that aims to decide between two possible distributions for the data, a hypothesis testing issue is investigated in this paper from a game theoretical point of view. More specifically: \textcolor{red}{\textbf{(\textit{i})}} To the best of our knowledge and from an information theoretic perspective, a MFG theoretical point of view has never ever been explored for the issues of hypothesis testing or/and \textit{maximum a posteriori} before. Therefore, it is revealed that the literature technically lacks in the theoretical context of the aforementioned issue $-$ something that is actualised in this paper. \textcolor{red}{\textbf{(\textit{ii})}} The Newton's second law, that is, $\mathbb{F}=ma$ is evaluated in relation to the inertia forces which characterise the information leakage to the network. \textcolor{red}{\textbf{(\textit{iii})}} In addition, some further discussions are gone over such as the rotation density in a given volume by calculating the curl of the leakage, or defining the suspension dynamics, stabilizability of the system is explored $-$ as well as the fact that the synchronisation of the system according to the asynchronous behaviour of the Byzantines are evaluated. 

\subsection{General notation \& preliminaries} 

Notations used throughout the paper are mainly given in Table \ref{table1}. The term $i \in \{ 1, \cdots, \mathscr{u} \}$ stands for the users where $\mathscr{U} \supseteq \mathscr{u}$.

\begin{table}[h]
\begin{center}
\captionof{table}{{List of useful notations.}}
 \label{table1} 
\begin{tabular}{||p{1.05cm}|p{3.28cm} |p{1.05cm}|p{1.46cm}||}
 \hline      
 \hline
Notation        &  Definition  & Notation          &  Definition   \\
\hline
       $\mathcal{W}$       &          Winner Walk       &    $t$        &       Time      \\
$(\cdot)^{(in)}$       &     Inner Game              &       $\theta$       &        Belief      \\
   $(\cdot)^{(out)}$        &         Outer Game           &    $(\cdot)^{\star}$        &        Optimum      \\
$\ell (\cdot|\cdot)$       &         Conditional Likelihood          &    $\mathscr{L}_{\alpha}$        &     $\alpha-$Leakage        \\
$\mathbb{E}$       &        Expected-value           &          $ \mathcal{X}$       &  State          \\
   $\mathcal{M}$        &      PDF of MFG       &    $\vartheta$       &         Cost       \\
   $\mathbb{N}ull$        &      Null       &    $\nabla$       &         Nabla      \\
   $Vol$        &      Volume      &    $\mathbb{B}\mathscr{all}$       &         Balloon     \\
\hline
 \hline
\end{tabular}
\end{center}
\end{table}

\subsection{Organisation}
The rest of the paper is organised as follows. The system set-up and our main results are given in Sections II and III. Subsequently, the evaluation of the framework and conclusions are given in Sections IV and V. The proofs are also presented in the appendices. The flow of the problem-and-solution in this paper is also shown in Fig. \ref{F0}. 

\section{System model and preliminaries}
In this section, we go over the opening topics. We initially discuss about the system model from both information theoretic and control theoretic standpoints and we also talk about some useful and fundamental preliminaries. 

Consider Fig. \ref{dool} $-$ temporarily considering only one Bob. Assume that a sender named Alice has some private data denoted by the random variable $\mathcal{S} \in S$ which is correlated with some non-private data $\mathcal{X} \in X$. Alice is supposed to share $\mathcal{X}$ with an analyst named Bob. However, due to the correlation between $\mathcal{X}$ and $\mathcal{S}$ which is captured by the joint distribution $\mathcal{P}_{\mathcal{S},\mathcal{X}}$, Bob may be able to draw some inference on the private data $\mathcal{S}$. Alice consequently decides to, instead of $\mathcal{X}$, release a distorted version of $\mathcal{X}$ defined by $\mathcal{Y} \in Y$ in order to alleviate the inference threat over $\mathcal{S}$ logically acquirable from the observation of $\mathcal{X}$. The distorted data $\mathcal{Y}$ is generated by passing through the following privacy mapping, i.e., the conditional distribution $\mathcal{P}_{\mathcal{Y}|\mathcal{X}}$. It should be noted that, in fact, Bob may also be able to act as an adversary by using $\mathcal{Y}$ to illegitimately infer the private data set $\mathcal{S}$, even though he is a legitimate recipient of the data set $\mathcal{Y}$. Therefore, the privacy mapping should be designed in the sense that we can be assured about a reduction to the inference threat on the private set $\mathcal{S}$ as follows: while preserving the utility of $\mathcal{Y}$ by maintaining the correlation, i.e., dependency between $\mathcal{Y}$ and $\mathcal{X}$, we aim at alleviating the dependency between $\mathcal{S}$ and $\mathcal{Y}$. This kind of two-fold information-theoretic goal balances a trade-off between utility and privacy. As also obvious, the Markov chain $\mathcal{S} \rightarrow \mathcal{X}\rightarrow \mathcal{Y}$ holds.

Let us go in datails from an information-theory point of view. Alice observes the source-symbol sets $-$ sequences $-$ $\mathcal{S}^{(\ell)}=\{ s^{(\ell)} \}^{\mathcal{L}_1}_{\ell=1}$ and $\mathcal{X}^{(\ell, q)}=\{ x^{(\ell, q)} \}^{\mathcal{L}_2}_{\ell=1}$ where $q-$th Bob observes $\mathcal{Y}^{(\ell, q)}=\{ y^{(\ell, q)} \}^{\mathcal{L}_3}_{\ell=1}$ which are i.i.d discrete memoryless variables respectively with the probability mass functions $\mathcal{P}_{\mathcal{S}}$, $\mathcal{P}_{\mathcal{X}}$ and $\mathcal{P}_{\mathcal{Y}}$ where $\{ \cdot \}^{\big(\ell \in \left \{  1,\cdots,\mathcal{L}_i, i \in \{ 1,2,3 \}  \right \}  \big)}$ stands literally for the $\ell$-th source-symbol per block belonging to the source-symbol set of size $1$-by-$\mathcal{L}_i, i \in \{ 1,2,3 \}$. In fact, we principally see a simple scheme where: (\textit{i}) the Alice-encoder follows the Borel measurable map $f^{(\ell, q)}_{x}: \;\mathcal{X}^{(\ell, q)}  \longmapsto \mathcal{M}^{(x, q)}=\{ 1, 2,\cdots, 2^{\ell \mathcal{R}^{(\ell, q)}_{x}} \}$, while $\sum{q} \mathcal{R}^{(\ell, q)}_{x}$ is the rate of Alice; moreover, (\textit{ii}) the Bob-decoder legitimately follows the Borel measurable map $g^{(\ell, q)}: \; \{ 1,\cdots, 2^{\ell \mathcal{R}^{(\ell, q)}_{x}} \}  \longmapsto \mathcal{X}^{(\ell, q)}$ after reception via the channel. The hypothesis is consequently defined as $\hat{\mathscr{H}} \triangleq g^{(\ell, q)} \big( \mathcal{M}^{(x, q)},   \mathcal{Y}^{(\ell, q)} \big)$ \cite{doool}.

Now, assume a finite set $\Theta \supseteq \theta$ of possible hypotheses where its cardinality declares the total number of hypotheses. In relation to the observation $\mathscr{s} \in \mathscr{S}$, the observation probability is given by a conditional likelihood function $\ell \big(\mathscr{s}|\theta^{\star} \big) \in [0;1]$, and $\sum\limits_{\mathscr{s} \in \mathscr{S}}\ell \big(\mathscr{s}|\theta^{\star}\big)= 1$. Meanwhile, the term $\theta^{\star}$ traditionally indicates the unknown but fixed true hypothesis to be learnt.

Additionally, for the $i \in \{1,\cdots,\mathscr{u} \}$ over the time horizon $t$, let us have the following essential definitions \cite{6, 12}.

\textsc{\textbf{Definition 1.}} Our basic definitions are as follows:
\begin{itemize}
\item \textsc{$\alpha-$leakage \cite{12}:} \textit{The $\alpha-$leakage\footnote{Relating to \textit{Sibson’s} mutual information: see e.g. \cite{fr}.} is defined as}
\begin{equation*}
\begin{split}
\; \mathscr{L}_{\alpha}\triangleq \frac{\alpha}{\alpha-1}log \frac{\mathop{{\rm max}}\limits_{\ell{\big(\theta^{\star}|\mathscr{s}} \big)}\mathbb{E}    \left \{ \Big( \ell{\big(\theta^{\star}=\theta|\theta,\mathscr{s}} \big) \Big )^{\frac{\alpha}{\alpha-1}}\right \}}{\mathop{{\rm max}}\limits_{\ell{\big(\theta^{\star}} \big)}\mathbb{E}    \left \{ \Big( \ell{\big(\theta^{\star}=\theta|\theta} \big) \Big )^{\frac{\alpha}{\alpha-1}}\right \}},
\end{split}
\end{equation*}
\textit{for $\alpha \in(1, \infty)$ and by the continuous extension for $\alpha =1, \infty$, without loss of generality;}
\item \textsc{Rate of convergence to true hypothesis \cite{6}:} \textit{Rate of convergence to true hypothesis $\theta^{\star}$ and over the time horizon is theoretically written by }
\begin{equation*}
\begin{split}
\; \omega^{(i)}_1 \triangleq \lim \limits_{t \rightarrow \infty }inf \bigg(    -\frac{1}{t} log \Big (\ell(\theta_i^{\star};i;t)-1 \Big) \bigg);
\end{split}
\end{equation*}
\item \textsc{Rate of convergence to false hypothesis \cite{6}:} \textit{Rate of convergence to true hypothesis $\theta^{\star}$ and over the time horizon is theoretically written by }
\begin{equation*}
\begin{split}
\; \omega^{(i)}_2 \triangleq \lim \limits_{t \rightarrow \infty }inf \bigg(    -\frac{1}{t} log \Big (\ell(\theta_i^{\star};i;t)-0 \Big) \bigg),
\end{split}
\end{equation*}
\item \textsc{Rate of social learning \cite{6}:} \textit{Rate of social learning and with regard to the true hypothesis $\theta^{\star}$ is presented as }
\begin{equation*}
\begin{split}
\; \omega_3 \triangleq \lim \limits_{t \rightarrow \infty }inf \bigg(    -\frac{1}{t} log \Big (\mathop{{\rm max}}\limits_{i} \sum \limits_{i=1}^{\mathscr{u}}\sum \limits_{j \neq i}\ell(\theta_j;t) \Big) \bigg).
\end{split}
\end{equation*}
\end{itemize}

\section{Main results}

Our main achievements are given in this section providing some novel propositions where the proofs are given in the appendices. 

\subsection{Objective $-$ \textit{''The Byzantine attack and defense are mutually exclusive but indivisible the same as spear and shield''}, declared in \cite{momi10}}

Defining the hypothesis pair $\big(  \mathscr{H}_0,\mathscr{H}_1  \big)$ as the following
\begin{equation*}
\begin{split}
\begin{cases}
\;  \mathscr{H}_1: \ell\big(\theta^{\star}\big)\ge \nu_1,\\
\mathscr{H}_0: \ell\big(\theta^{\star}\big)< \nu_1,
\end{cases}
\end{split}
\end{equation*}
w.r.t. the threshold $\nu_1$, the following error probability $\mathbb{P}_{\mathscr{e}}$ should be minimised 
\begin{equation*}
\begin{split}
\; \mathbb{P}_{\mathscr{e}} \triangleq \mathbb{P} \Big (  \mathscr{H}_1 \big| \mathscr{H}_0  \Big)+\mathbb{P} \Big (  \mathscr{H}_0 \big| \mathscr{H}_1  \Big),
\end{split}
\end{equation*}
where the first term is called the false alarm probability, and the second one stands for the miss detection probability. Indeed, Byzantines aim to the cooperation in the network to fail via an increment in both the false alarm probability and the miss detection probability. They do this in the sense that they can conveniently do their own communications in a more efficient manner at the cost of other users' loss\footnote{Byzantine naturally arise when the resources in the overall network is scarce.}. The Byzantine attack issue principally is in fact a zero-sum game between the detection process and malicious users and that is additionally a pursuit-evation game for Byzantines. 

\textsc{\textbf{Remark 0.}} \textit{We use the temporal rate of the $\alpha-$leakage as the tool which measures the resilience.}

Now, let us go in details, where in Fig. \ref{F0} the flow of the problem-and-solution is briefly shown.

\subsection{$\mathbb{F}=ma$}
\begin{proposition} \label{P2} \textit{The inertia forces which actualise the information-leakage are attainable.}\end{proposition}

\textbf{\textsc{Proof:}} See Appendix \ref{sec:A}.$\; \; \; \blacksquare$

\begin{corollary}{According to the third law of thermodynamics, when the temporal rate of the $\alpha-$leakage goes to be constant if and only if the temperature set, that is, the term $\theta$ thechnically goes to zero.}
\end{corollary}

\textbf{\textsc{Proof:}} The proof is easy to follow, thus, it is neglected here.$\; \; \; \blacksquare$

\subsection{MFG}
\begin{proposition} \label{P2} \textit{In the consequence of Proposition 1 since one consider it as the control law of an MFG, there exists a couple of HJB and FPK equations, which can thoroughly solve the game.}\end{proposition}

\textbf{\textsc{Proof:}} See Appendix \ref{sec:B} $-$ Similar to \cite{444}.$\; \; \; \blacksquare$

\subsection{Stabilizability}
\begin{proposition} \label{P2} \textit{The overall system may experience an absolute stabilizability.}\end{proposition}

\textbf{\textsc{Proof:}} See Appendix \ref{sec:C}.$\; \; \; \blacksquare$

\subsection{Nash-equilibrium}

\begin{proposition}{Our bi-level MFG even experiences a Nash-equilibrium where its probability is bounded.}\end{proposition}

\textbf{\textsc{Proof:}} See Appendix \ref{sec:D}.$\; \; \; \blacksquare$

\begin{algorithm}
\caption{\textcolor{black}{A \textit{Dog-Leg}-method based ADMM-oriented iterative online algorithm for computing $\frac{d\ell  \big ( \theta | \mathscr{s}_0 \big )}{d\ell  \big ( \theta  \big )}$ named \textsc{$\mathbb{F}$unction $dog-leg$}.}}\label{ffff}
\begin{algorithmic}
\STATE \textbf{\textsc{$\mathbb{F}$unction}} $dog-leg$

\STATE \textbf{\textsc{Input:}} $\theta  $, $ \Theta \supseteq \theta  $, $\mathscr{u}, \;\mathscr{U} \supseteq \mathscr{u}$, $d\ell  \big ( \theta | \mathscr{s}_0 \big ) $, $d\ell  \big ( \theta  \big ) $, $\mathscr{s}_0 $.

\STATE \textbf{\textsc{Initialisation:}} {Set} $\frac{d\ell  \big ( \theta | \mathscr{s}_0 \big )}{d\ell  \big ( \theta  \big )}$ and the input parameters as $\emptyset$.

$\;$\textbf{for} $i \in \mathscr{U}$ \textbf{do}

$\;\;\;$\textbf{for} $k-$th iteration \textbf{do}

$\;\;\;\;\;\;\;\;\;\;\;\;\;\;\;\;\;\;\;\;$ \textbf{while} $\bigg|  \frac{d\ell_{k,i}  \big ( \theta | \mathscr{s}_0 \big )}{d\ell_{k,i}  \big ( \theta  \big )}       -   \frac{d\ell_{k,i}  \big ( \theta | \mathscr{s}_0 \big )}{d\ell_{k-1,i}  \big ( \theta  \big )}    \bigg  | > \phi_{k,i}$ \textbf{do}

$\;\;$ (\textit{a.i.}) Set $d\ell  \big ( \theta  \big )$ to be fixed as $d\ell^{{f}}  \big ( \theta  \big )$; 

$\;\;$ (\textit{a.ii.}) Solve $\mathop{{\rm argmin}}\limits_{\theta} \frac{d\ell  \big ( \theta | \mathscr{s}_0 \big )}{d\ell^{{f}}  \big ( \theta  \big )}$; and

$\;\;$ (\textit{a.iii.}) Update: $ \Big( \theta^{\star} ;d\ell^{\star}  \big ( \theta^{\star} | \mathscr{s}_0 \big ) \Big)\gets   \Big( \theta_{k,i} ;d\ell_{k,i}   \big ( \theta_{k,i}  | \mathscr{s}_0 \big ) \Big)  $.

$\;\;$ (\textit{b.i.}) Set $d\ell \big ( \theta | \mathscr{s}_0 \big )$ to be fixed as $d\ell^{{f}}  \big ( \theta | \mathscr{s}_0 \big )$; 

$\;\;$ (\textit{b.ii.}) Solve $\mathop{{\rm argmin}}\limits_{\theta} \frac{d\ell^{{f}}  \big ( \theta | \mathscr{s}_0 \big )}{d\ell  \big ( \theta  \big )}$; and

$\;\;$ (\textit{b.iii.}) Update: $ \Big( \theta^{\star} ;d\ell^{\star}  \big ( \theta^{\star}  \big ) \Big)\gets   \Big( \theta_{k,i} ;d\ell_{k,i}   \big ( \theta_{k,i}   \big ) \Big)  $.

$\;\;\;\;\;\;\;\;\;\;\;\;\;\;\;\;\;\;\;\;$ \textbf{endwhile} 

$\;\;\;\;\;$ \textbf{endswitch}

$\;\;\;\;\;k \gets k+1$

$\;\;\;$ \textbf{endfor}

$\;\;\;i \gets i+1$

$\;$ \textbf{endfor}

\textbf{\textsc{Return}} $\frac{d\ell^{\star}  \big ( \theta^{\star} | \mathscr{s}_0 \big )}{d\ell^{\star}  \big ( \theta^{\star}  \big )}$

\STATE \textsc{\textbf{END-function}}

\end{algorithmic}
\end{algorithm}

\subsection{Vector field}
\begin{proposition}{$\alpha-$leakage $\mathscr{L}_{\alpha}$ is a vector field through which the rotation density in each point over the given surfaces and enclosed loops 1-dimensionally or/and 2-dimensionally created by the likelihood functions $\ell(\cdot)$ is analysable $-$ whether or not this density is equated with zero.}\end{proposition}

\textbf{\textsc{Proof:}} See Appendix \ref{sec:E}.$\; \; \; \blacksquare$

\begin{corollary}{According to the third law of thermodynamics and Corollary 1, the $\alpha-$leakage is not curl-free, although according to Proposition 4 it can be, relating to the Nash-equilibrium.}
\end{corollary}

\textbf{\textsc{Proof:}} The proof is easy to follow, thus, it is neglected here $-$ the divergence is non-conservative over the closed-loops enclosed to the change in the likelihood functions, i.e., $d \ell(\cdot)$.$\; \; \; \blacksquare$

\subsection{Ascending/descending nature }
\begin{proposition}{So far, it has been proven that the $\ell(\cdot)$ terms are linear or non-linear functions of each other $-$ in the appendices. Now, one can assert that for each group of Byzantines and good agents separately, all the $\ell(\cdot)$ are of either a purely ascending nature or a descending one.}\end{proposition}

\textbf{\textsc{Proof:}} See Appendix \ref{sec:F}.$\; \; \; \blacksquare$

\subsection{Probabilistic formalisation of the true hypothesis $\mathscr{H}_1$}
\begin{proposition}{The true hypothesis $\mathscr{H}_1$ can be mapped to a probabilistic one where it is both attainable and analysable. }\end{proposition}

\textbf{\textsc{Proof:}} See Appendix \ref{sec:G}.$\; \; \; \blacksquare$

\begin{algorithm*}
\caption{\textcolor{black}{An iterative online algorithm for byzantine-resilient distributed hypothesis testing named $\mathbb{OBRDHT}$.}}\label{fff}
\begin{algorithmic}
\STATE \textbf{\textsc{$\mathbb{P}$rocedure}} $\mathbb{OBRDHT}$

\STATE \textbf{\textsc{Input:}} 

$\;\;\;\;\mathbb{I}\text{nput} \triangleq \bigg \lbrace \theta, \Theta \supseteq \theta, \mathscr{u},\mathscr{U} \supseteq \mathscr{u}, \mathcal{T}^{(case)}_{start}$,
$\mathcal{T}^{(case)}_{end}$,
$\mathcal{X}^{(case)}_i$,
$\mathcal{X}^{(case)}_{-i}$,
$c^{(case)}_{1}$, $c^{(case)}_{2}$, $c^{(case)}_{3}$, $\zeta^{(case)}_{local} (\cdot)$,
$\mathcal{M}^{(case)}$, $\frac{d\ell  \big ( \theta | \mathscr{s}_0 \big )}{d\ell  \big ( \theta  \big )} \bigg \rbrace$

\STATE \textbf{\textsc{Initialisation:}} 

$\;\;\;\;\;\;\;\;\;\;\;\;\;\;\;\;\;\;\;\;\;\;\;\;\;\;\;$ (\textit{i}) Invoke $\frac{d\ell  \big ( \theta | \mathscr{s}_0 \big )}{d\ell  \big ( \theta  \big )}$ from $\mathbb{F}$unction $dog-leg$.  ${\;\%\; Algorithm \; \ref{ffff}} \cdots$

$\;\;\;\;\;\;\;\;\;\;\;\;\;\;\;\;\;\;\;\;\;\;\;\;\;\;\;$ (\textit{ii}) {Set} $\ell \big ( \theta \big) $ and $\bigg \lbrace\mathbb{I}\text{nput} \setminus \bigg \lbrace\frac{d\ell  \big ( \theta | \mathscr{s}_0 \big )}{d\ell  \big ( \theta  \big )}\bigg \rbrace \bigg \rbrace$ as $\emptyset$.

$\;$\textbf{for} $i \in \mathscr{U}$ \textbf{do}

$\;\;\;$\textbf{for} $k-$th iteration \textbf{do}

$\;\;\;\;\;$\textbf{switch} $case$ $\;\;\;\;\;\; \;\%\; (\cdot)^{(case)} \in \{ (\cdot)^{(in)},(\cdot)^{(out) } \}$

$\;\;\;\;\;\;\;\;\;$\textbf{case} 1

$\;\;\;\;\;\;\;\;\;\;\;\;\;\;\;\;\;$ $case = in $ $ \;\;\;\;\;\;\;\%\; (\cdot)^{(case)} = (\cdot)^{(in)}$

$\;\;\;\;\;\;\;\;\;$\textbf{case} 2

$\;\;\;\;\;\;\;\;\;\;\;\;\;\;\;\;\;$ $case = out $ $\;\;\; \;\;\;\;\%\; (\cdot)^{(case)} = (\cdot)^{(out) } $

$\;\;\;\;\;\;\;\;\;\;\;\;\;$\textbf{if} $\theta \neq \emptyset \; \textcolor{red}{\&} \; \Theta  \neq \emptyset \; \textcolor{red}{\&}\;  \mathscr{u} \neq \emptyset \; \textcolor{red}{\&} \; \mathscr{U}  \neq \emptyset \; \textcolor{red}{\&}\;  \mathcal{T}^{(case)}_{start} \neq \emptyset \; \textcolor{red}{\&}\; 
\mathcal{T}^{(case)}_{end} \neq \emptyset \; \textcolor{red}{\&} \; 
\mathcal{X}^{(case)}_i \neq \emptyset \; \textcolor{red}{\&}\; 
\mathcal{X}^{(case)}_{-i} \neq \emptyset \; \textcolor{red}{\&}\; 
c^{(case)}_{1} \neq \emptyset \; \textcolor{red}{\&}\; c^{(case)}_{2} \neq \emptyset \; \textcolor{red}{\&}\; c^{(case)}_{3} \neq \emptyset \; \textcolor{red}{\&} \; \zeta^{(case)}_{local} (\cdot) \neq \emptyset \; \textcolor{red}{\&}\; 
\mathcal{M}^{(case)} \neq \emptyset \; \textcolor{red}{\&} \; \frac{d\ell  \big ( \theta | \mathscr{s}_0 \big )}{d\ell  \big ( \theta  \big )} $ \textbf{then}

$\;\;\;\;\;\;\;\;\;\;\;\;\;\;\;\;\;\;\;\;$ \textbf{while} $ \Big|\mathcal{X}^{(case)}_{k,i}-\mathcal{X}^{(case)}_{k-1,i}  \Big| > \gamma_{k,i}^{(case)}$ \textbf{do}

$\;\;\;\;\;\;\;\;\;\;\;\;\;\;\;\;\;\;\;\;\;\;\;\;\;\;\;\;\;\;\;$ \textsc{Phase 0: at Server.} Server sends a predefined global belief to Users.

$\;\;\;\;\;\;\;\;\;\;\;\;\;\;\;\;\;\;\;\;\;\;\;\;\;\;\;\;\;\;\;$ \textsc{Phase 1: at Users.} Users send their local beliefs to Server.

$\;\;\;\;\;\;\;\;\;\;\;\;\;\;\;\;\;\;\;\;\;\;\;\;\;\;\;\;\;\;$ \textsc{Phase 2: at Server.} Server receives the local observations from Users and compares the gradient of the local beliefs of Users with the global belief according to the relative control laws: 

$\;\;\;\;\;\;\;\;\;\;\;\;\;\;\;\;\;\;\;\;\;\;\;\;\;\;\;\;\;\;\;\;\;\;\;\;\;\;\;\;\;\;\;\;\;\;\;\;\;\;\;\;\;\;\;\;\;\;\;\;\;\;\;\;\;\;\;\;\;\;\;\;$ (\textit{i}) Find $\ell_{k,i} \big ( \theta \big)$ w.r.t. HJB-FPK; 

$\;\;\;\;\;\;\;\;\;\;\;\;\;\;\;\;\;\;\;\;\;\;\;\;\;\;\;\;\;\;\;\;\;\;\;\;\;\;\;\;\;\;\;\;\;\;\;\;\;\;\;\;\;\;\;\;\;\;\;\;\;\;\;\;\;\;\;\;\;\;\;\;$ ${\;\%\; Record \;the\; optimum} \cdots$

$\;\;\;\;\;\;\;\;\;\;\;\;\;\;\;\;\;\;\;\;\;\;\;\;\;\;\;\;\;\;\;\;\;\;\;\;\;\;\;\;\;\;\;\;\;\;\;\;\;\;\;\;\;\;\;\;\;\;\;\;\;\;\;\;\;\;\;\;\;\;\;\;$ (\textit{ii}) Update: $   \ell \big ( \theta^{\star} \big)     \gets  \ell_{k,i} \big ( \theta \big)$; and

$\;\;\;\;\;\;\;\;\;\;\;\;\;\;\;\;\;\;\;\;\;\;\;\;\;\;\;\;\;\;\;\;\;\;\;\;\;\;\;\;\;\;\;\;\;\;\;\;\;\;\;\;\;\;\;\;\;\;\;\;\;\;\;\;\;\;\;\;\;\;\;\;$ (\textit{iii}) Server sends the global belief to Users.

$\;\;\;\;\;\;\;\;\;\;\;\;\;\;\;\;\;\;\;\;$ \textbf{endwhile} 

$\;\;\;\;\;\;\;\;\;\;\;\;\;$\textbf{else} 

$\;\;\;\;\;\;\;\;\;\;\;\;\;$\textbf{then} Break;

$\;\;\;\;\;\;\;\;\;\;\;\;\;$\textbf{endif} 

$\;\;\;\;\;$ \textbf{endswitch}

$\;\;\;\;\;k \gets k+1$

$\;\;\;$ \textbf{endfor}

$\;\;\;i \gets i+1$

$\;$ \textbf{endfor}

\textbf{\textsc{Return}} $\ell \big ( \theta^{\star} \big)$

\STATE \textsc{\textbf{END-procedure}}

\end{algorithmic}
\end{algorithm*}

\subsection{How to deal with the asynchronous behaviour of the Byzantines}

There arises a tremendously crucial query that what if the Byzantines asynchronise themselves and if there exists a solution to this type of issue. The greedy Algorithm \ref{fo} as well as Algorithm \ref{foo} indeed provide an opportunity for us to tackle this kind of issue.
\begin{proposition}{One can probabilistically go towards a solution to the asynchronous behaviour of the Byzantines.}\end{proposition}

\textbf{\textsc{Proof:}} See Appendix \ref{sec:H}.$\; \; \; \blacksquare$

\begin{proposition}{One can go towards a solution to the sasynchronous behaviour of the Byzantines from a non-convex optimisation point of view.}\end{proposition}

\textbf{\textsc{Proof:}} See Appendix \ref{sec:I}.$\; \; \; \blacksquare$

\begin{proposition}{The criterion in relation to the freshness of the information, that is, the age-of-information (AoI) is analysable in a given domain.}\end{proposition}

\textbf{\textsc{Proof:}} See Appendix \ref{sec:J}.$\; \; \; \blacksquare$

\subsection{Complexity}

The online\footnote{An online algorithm is the one where the data set is only available in terms of a real-time fashion, in contrast to offline ones where the whole data set is completely available at the initial point.} Algorithm \ref{fff} named $\mathbb{OBRDHT}$ has the following complexity. In each iteration and for every user, our MFG principally creates a complexity order of bound $NlogN$, where the function $\mathbb{F}\textit{unction} \;dog-leg$ being invoked from Algorithm \ref{ffff} including the alternating direction method of multipliers (ADMM) creates the one as $\mathcal{O}\big(\rho^{-1}\big)+\mathcal{O}\big(\rho^{-2}\big)$ w.r.t. the given accuracy coefficient $\rho$. Therefore, the total mount of the complexity has the order of bound $\mathcal{T}\times\mathscr{u} \times \Big(\mathcal{O}\big(\rho^{-1}\big)+\mathcal{O}\big(\rho^{-2}\big)+NlogN \Big)$ where $\mathcal{T}$ stands literally for the overall running time.

\section{Numerical results}

We have done our simulations w.r.t. the Bernoulli-distributed data-sets using GNU Octave of version $4.2.2$ on Ubuntu $16.04$. The parameters used in the simulations are listed in Table \ref{table2} \cite{777, 666}. Recall $\mathbb{P}_{\mathscr{e}} \triangleq \mathbb{P} \Big (  \mathscr{H}_1 \big| \mathscr{H}_0  \Big)+\mathbb{P} \Big (  \mathscr{H}_0 \big| \mathscr{H}_1  \Big)$ from the previous parts while the first term is the false alarm probability as $\frac{\kappa_2}{\kappa_1+\kappa_2}\mathscr{Q} \Big(  \frac{\partial-2\tau_s B_a \partial_{n}^2}{\sqrt{4\tau_s B_a \partial_{n}^4}}  \Big)$, and the second one stands for the miss detection probability as $\frac{\kappa_1}{\kappa_1+\kappa_2}\bigg(1-\mathscr{Q} \Big(  \frac{\partial-2\tau_s B_a (\partial_{p}^2+\partial_{n}^2)}{\sqrt{4\tau_s B_a (\partial_{p}^2+\partial_{n}^2)^2}}  \Big) \bigg)$ while $\partial$ is the detection threshold, $\mathscr{Q}(\cdot)$ is Q-function and $\tau_s$ is the sensing time while $\tau_s=\tau_{tot}- \tau_t$ holds w.r.t. the transmission time $\tau_t$ calculable as $\tau_t=\frac{E_t}{\frac{\kappa_4 N_{0}N_{rx}B_a (\frac{4 \pi}{\kappa_5})^{\kappa_3}10^{\kappa_3}}{G_a\kappa_6}r_s^{\kappa_3}+P_{elec}}$ w.r.t. the energy consumed for transmission  $E_t$ and the transmission range $r_s$. It is noteworthy to add that we get access to the energy from the MFG theoretical discussion and the Hamiltonian, and the time frame w.r.t. AoI. 

\begin{table}[h]
\begin{center}
\captionof{table}{{Simulation parameters $-$ similar to \cite{777, 666}.}}
 \label{table2} 
\begin{tabular}{||p{4.25cm}|p{0.88cm} |p{1.64cm}||}
 \hline      
 \hline
Parameter        &  Symbol     &  Value  \\
\hline
PU birth rate& $\kappa_1$&$3 / s$ \\
PU death rate& $\kappa_2$& $3 / s$ \\
Path loss exponent &$\kappa_3$&$ 3$ \\
Reference SU SNR (dB)&$\kappa_4$& $ 20$ \\
Byzantine-PU SNR (dB)& $\frac{\partial^2_p}{\partial^2_n}$& $10  \frac{d\ell  \big ( \theta | \mathscr{s}_0 \big )}{d\ell  \big ( \theta  \big )} $ \\
Non-Byzantine-PU SNR (dB)& $\frac{\partial^2_p}{\partial^2_n}$& $10  $ \\
Receiver noise figure &$N_{rx}$& $12.589$ \\
Thermal noise ($\frac{W}{Hz}$) &$N_0$ &$417 \times 10^{-23} $ \\
Bandwidth &$B_a $&$10 KHz$ \\
Signal wavelength &$\kappa_5$&$ 0.125 m$ \\
Amplifier efficiency &$\kappa_6 $&$0.2$ \\
Antenna gain &$G_a$& $0.01$ \\
TX circuit power cons. &$P_{elec}$&$3.63 mW$ \\
\hline
 \hline
\end{tabular}
\end{center}
\end{table}

\begin{figure}[t]
\centering
\subfloat{\includegraphics[trim={{17 mm} {64 mm} {21 mm} {74mm}},clip,scale=0.4]{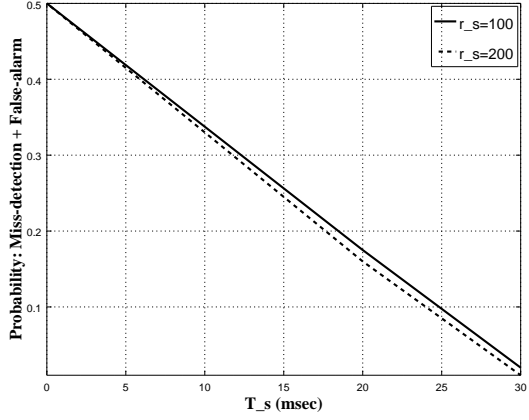}} 
\caption{total error $\mathbb{P}_{\mathscr{e}} \triangleq \mathbb{P} \Big (  \mathscr{H}_1 \big| \mathscr{H}_0  \Big)+\mathbb{P} \Big (  \mathscr{H}_0 \big| \mathscr{H}_1  \Big)$ as discussed in the previous paragraph is depicted versus the $\tau_s$ regime while changing $r_s$. } \label{Fm}
\end{figure}

In Fig. \ref{Fm}, the total error $\mathbb{P}_{\mathscr{e}} \triangleq \mathbb{P} \Big (  \mathscr{H}_1 \big| \mathscr{H}_0  \Big)+\mathbb{P} \Big (  \mathscr{H}_0 \big| \mathscr{H}_1  \Big)$ as discussed in the previous paragraph is depicted versus the $\tau_s$ regime while changing $r_s$. The more $r_s$ we choose, the less probability we can assure. 

\begin{figure}[t]
\centering
\subfloat{\includegraphics[trim={{17 mm} {64 mm} {21 mm} {74mm}},clip,scale=0.4]{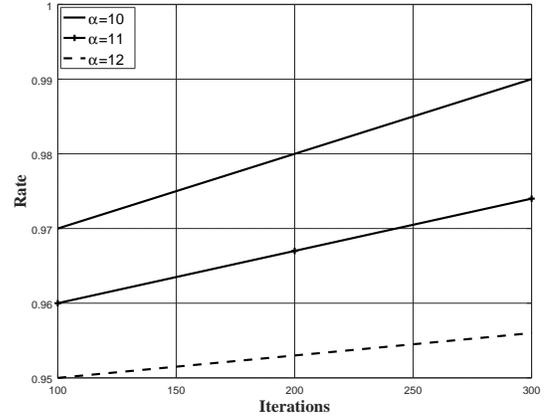}} 
\caption{Rate including $\alpha-$leakage against the iteration regime, changing $\alpha$. } \label{F1}
\label{fig:EcUND} 
\end{figure}

In Fig. \ref{F1}, the $\alpha-$included rate versus the iteration set is depicted. As obvious, the higher $\alpha$ we have, the worse overall system performance we experience.

\begin{figure}[t]
\centering
\subfloat{\includegraphics[trim={{17 mm} {64 mm} {21 mm} {74mm}},clip,scale=0.4]{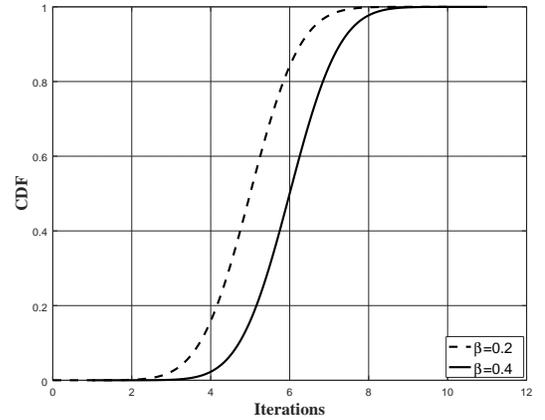}} 
\caption{CDF of the iterations required for the convergence of online Algorithm \ref{fff} $-$ named $\mathbb{OBRDHT}$ $-$ changing the normalised version of $\frac{d\ell  \big ( \theta | \mathscr{s}_0 \big )}{d\ell  \big ( \theta  \big )}$, i.e., $\beta$. } \label{F2}
\label{fig:EcUND} 
\end{figure}

In Fig. \ref{F2}, the cumulative distribution function (CDF) of the iterations required for the convergence of Algorithm \ref{fff} named $\mathbb{OBRDHT}$ $-$ which invokes Algorithm \ref{ffff} named $dog-leg$ $-$ is presented while changing the normalised version of $\beta \triangleq \frac{d\ell  \big ( \theta | \mathscr{s}_0 \big )}{d\ell  \big ( \theta  \big )}$. It is revealed that our hidden control law which will be discussed in Appendix \ref{sec:B} $-$ Remark 4 $-$ is a highly important key in our scheme $-$ something that emphatically proves our contribution in this context. 

\begin{figure}[t]
\centering
\subfloat{\includegraphics[trim={{17 mm} {64 mm} {21 mm} {74mm}},clip,scale=0.4]{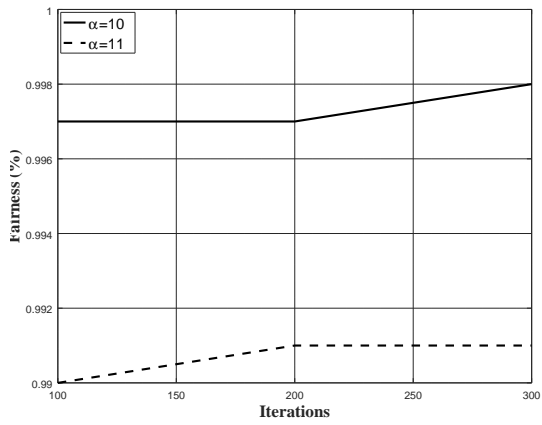}} 
\caption{Fairness index (\%) versus the number of the iterations changing $\alpha$.} \label{F3}
\label{fig:EcUND} 
\end{figure}

In Fig. \ref{F3}, in terms of the Jain's fairness index, the fairness
\begin{equation*}
\begin{split}
\; \frac{\bigg(\sum\limits_{i=1}^u \mathscr{h}\Big(\big(  \mathscr{L}^{(i)}_{\alpha} \big)^{-1}\Big)\bigg)^2}{u \sum\limits_{i=1}^u \mathscr{h}\Big(\big(  \mathscr{L}^{(i)}_{\alpha} \big)^{-1}\Big)^2},
\end{split}
\end{equation*}
is shown while changing $\alpha$ where $\mathscr{h} \big( \cdot \big)$ is a reverse function of $\mathscr{L}^{(i)}_{\alpha} $. As absolutely obvious, the same as Fig. \ref{F1}, the higher $\alpha$ we have, the worse overall system performance we literally experience.

\begin{figure}[t]
\centering
\subfloat[$\alpha=8$]{\includegraphics[trim={{17 mm} {64 mm} {21 mm} {74mm}},clip,scale=0.4]{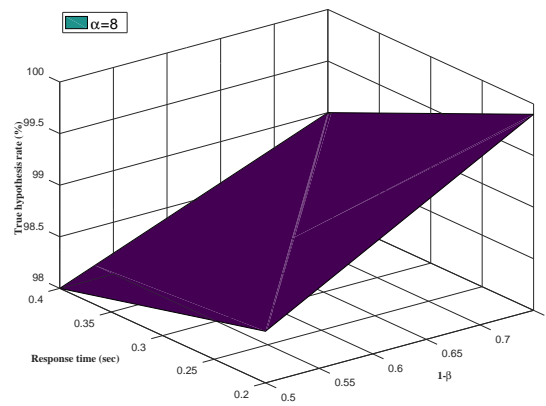}} \\
\subfloat[$\alpha=18$]{\includegraphics[trim={{17 mm} {64 mm} {21 mm} {74mm}},clip,scale=0.4]{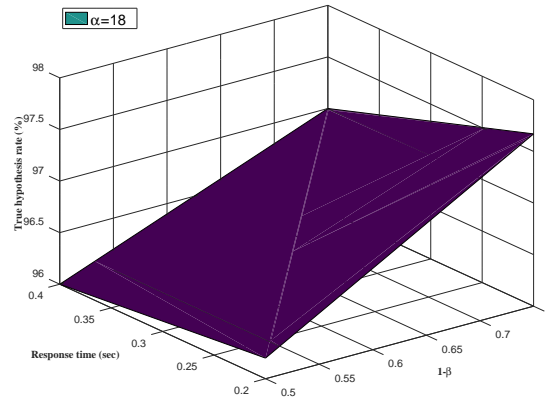}} 
\caption{Tru hypothesis rate(\%) versus the response time (second) vs. the $1-\beta$ regime, changing $\alpha$ $-$ something that recoreds different maximum values.} \label{F5}
\label{fig:EcUND} 
\end{figure}

In Fig. \ref{F5}, the true hypothesis rate (\%), that is, $\omega_1$ from Definition 1 is shown versus the response time vs. the $1-\beta$ regime, while changing $\alpha$. This figure also emphasises on the results achieved from the previous figures $-$ something that recoreds different maximum values.

\begin{figure}[t]
\centering
\subfloat[$\alpha-$leakage obtained from Algorithm \ref{fo} divided to the one obtained by Algorithm \ref{foo}.]{\includegraphics[trim={{14 mm} {64 mm} {21 mm} {74mm}},clip,scale=0.4]{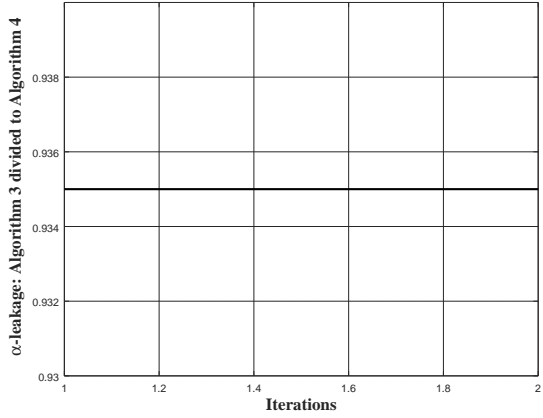}}\\
\subfloat[CDF of synchronisation vs. the iterations while changing $\alpha$ and $\beta$: Algorithm \ref{foo}.]{\includegraphics[trim={{14 mm} {64 mm} {21 mm} {74mm}},clip,scale=0.4]{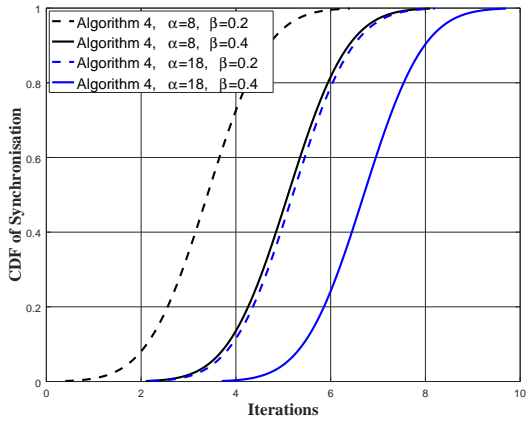}}\\
\subfloat[CDF of synchronisation vs. the iterations while changing $\alpha$ and $\beta$: Algorithm \ref{fo}.]{\includegraphics[trim={{14 mm} {64 mm} {21 mm} {74mm}},clip,scale=0.4]{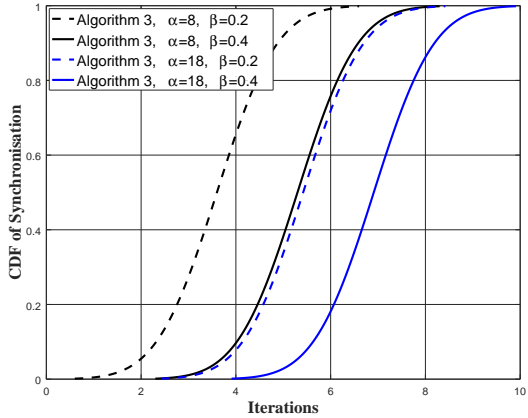}}
\caption{An overall comparison between Algorithm \ref{fo} and Algorithm \ref{foo}.} \label{F6}
\label{fig:EcUND} 
\end{figure}

In Fig. \ref{F6}, pen-ultimately speaking, Algorithm \ref{fo} and Algorithm \ref{foo} are compared with each other. The first sub-figure shows the amount of $\alpha-$leakage obtained from Algorithm \ref{fo} divided to the one obtained by Algorithm \ref{foo} is shown. As revealed, Algorithm \ref{foo} out-performs Algorithm \ref{fo} in the context of synchronisation. Meanwhile, the second and the third sub-figures also show a significantly more adequate and acceptable CDF of synchronisation for Algorithm \ref{foo} compared to Algorithm \ref{fo} $-$ for the various values of $\alpha$ and $\beta$.

\begin{figure}[t]
\centering
\subfloat[Average optimal policy $\pi_{t}(a|s)$]{\includegraphics[trim={{17 mm} {64 mm} {21 mm} {74mm}},clip,scale=0.4]{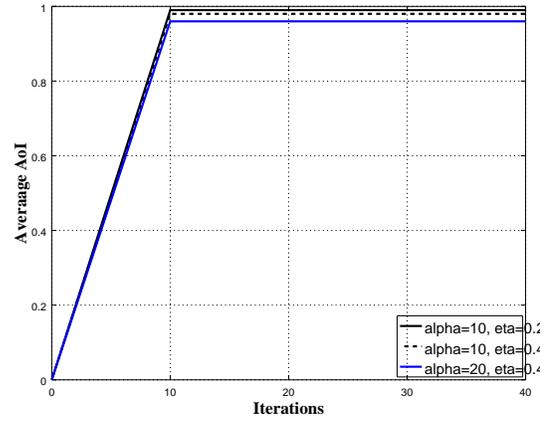}} 
\caption{Average normalised AoI versus the iterations while changing $\alpha$ and $\beta$..} \label{F11}
\label{fig:EcUND} 
\end{figure}

In Fig. \ref{F11}, finally speaking, the average normalised AoI versus the iteration regime is finally depicted while changing the values for $\alpha$ and $\beta$. This figure also, as obvious, proves the effect of $\alpha$ and $\beta$ the same as the previous figures as well. 

\section{conclusion}
Novel insights to byzantine-resilient distributed hypothesis testing were provided in this paper. We proposed a new iterative algorithm and examined its performance by simulations and interpreted it from a theoretical point of view as well.

\appendices
\section{Proof of Proposition 1: \\$\mathbb{F}=ma$}
\label{sec:A}
For the proof, we need to go over the Newton's second law as $\mathbb{F}={ma}$ as below.

Initially speaking, the temporal rate of the $\alpha-$leakage can be principally calculated as
\begin{equation*}
\begin{split}
\; \frac{d \mathscr{L}_{\alpha}}{dt} \triangleq \;\;\;\;\;\;\;\;\;\;\;\;\;\;\;\;\;\;\;\;\;\;\;\;\;\;\;\;\;\;\;\;\;\;\;\;\;\;\;\;\;\;\;\;\;\;\;\;\;\;\;\;\;\;\;\;\;\;\;\;\;\;\;\;\;\;\;\;\;\;\;\;\;\;\;\;\;\; \\ \underbrace{\frac{d \mathscr{L}_{\alpha}}{d \ell \big ( \theta^{\star}| \mathscr{s} \big ) } \times  \cancelto{\ell  \big (  \theta^{\star} | \theta \big )}{\frac{d \ell  \big (  \theta^{\star} | \mathscr{s} \big ) }{d \ell  \big ( \theta | \mathscr{s} \big ) } }   \times \frac{d \ell  \big ( \theta | \mathscr{s}\big )}{d \ell \big ( \theta | \mathscr{s}_0 \big )}\times  \frac{d\ell  \big ( \theta | \mathscr{s}_0 \big )}{d\ell  \big ( \theta  \big )}}_{\text{Outer MFG}}\times \underbrace{\frac{d\ell  \big ( \theta  \big )}{dt}}_{\text{Inner MFG}},
\end{split}
\end{equation*}
according to the chain rule in derivatives, where: 
\begin{itemize}
\item (\textit{i}) the first term comes from the fact that $\mathscr{L}_{\alpha}$ is mainly a function of $\ell \big ( \theta^{\star}| \mathscr{s} \big )$; 
\item (\textit{ii}) for the second term, that is, $\frac{d \ell  \big (  \theta^{\star} | \mathscr{s} \big ) }{d \ell  \big ( \theta | \mathscr{s} \big ) }$ which is equal to $\ell  \big (  \theta^{\star}| \theta\big )$, we need\footnote{See e.g. \cite{13}.} $\ell  \big ( {\mathscr{s}} |  \theta^{\star} \big )$ and $\ell \big ( {\mathscr{s}} | \theta \big )$ which are both available; 
\item (\textit{iii}) the third term theoretically originates from the fact that we need to define a  constraint according to which the divergence between the current state with the major player $-$ in terms of a hidden game where the observations are the agents $-$ , that is, the initial condition can be less than a threshold\footnote{See e.g. \cite{11}.}; 
\item (\textit{iv}) the penultimate term comes from the issue that we are interesting in doing a distillation of $\ell  \big ( \theta | \mathscr{s}_0 \big )$ in finding $\ell  \big ( \theta  \big )$ $-$ as it is possible since $\ell  \big ( \theta  \big )=\sum\limits_{\mathscr{s}_0 }\ell  \big ( \mathscr{s}_0 \big )\ell  \big ( \theta | \mathscr{s}_0 \big )$ holds; and 
\item (\textit{v}) the last term can be justified by the \textit{Kushner-Zakai} principle\footnote{See e.g. \cite{14}.}.
\end{itemize}

Regarding the issue that all the four terms represented above are achievable, therefore, we can say that the temporal rate of the $\alpha-$leakage is attainable as well.

\textsc{\textbf{Remark 1.}} \textit{The iterative Dog-leg-method based ADMM oriented online algorithm \ref{ffff} is presented for the issue expressed here. This algorithm can find the optimum value for the pair $\Big( d\ell^{\star} \big ( \theta^{\star} | \mathscr{s}_0 \big ) ; d\ell^{\star}  \big ( \theta^{\star}  \big ) \Big)$ as the Leg-Knee pair.}

The proof is now completed.$\; \; \; \blacksquare$

\section{Proof of Proposition 2:\\ MFG realisation $-$ Similar to \cite{444}}
\label{sec:B}
The sketch of the proof is easy to follow. First, we consider a nested bi-level MFG for which we define $(\cdot)^{(in)}$ and $(\cdot)^{(out)}$, respectively. Indeed, there exists\footnote{See e.g. \cite{16}.} a bi-level online stochastic optimisation as $
\; \mathop{{\rm min}}\limits_{\ell\big ( \theta\big)} \mathbb{E}_{\theta} \bigg( \Omega_1 \Big(\ell\big ( \theta\big), \mathscr{L}_{\alpha} \Big) \Big|\mathscr{H}_1 \bigg)$$ s.t.\;\ \mathop{{\rm min}}\limits_{\theta} \mathbb{E}_{t} \Big(\Omega_2 \big(  \theta,t  \big) \Big |\mathscr{H}_1\Big),
$
w.r.t. the given $\Omega_1(\cdot)$ and $\Omega_2 (\cdot)$.\footnote{\textsc{Step 1: Control law.}
\begin{itemize}
\item \textcolor{black}{{(\textit{i}) Inner MFG.}} Defining the state $\mathcal{X}^{(in)}_i(t) \triangleq \ell ^{(i,t)}\big ( \theta \big )$ in relation to the $i-$th agent, we have the control law as the game dynamics as follows $d \mathcal{X}^{(in)}_i(t) = \mathscr{f}^{(in)}\big ( \mathcal{X}^{(in)}_i(t)  \big ) dt+ \mathcal{W}^{(in)}_{i,0}$ $-$ as the \textit{Kushner-Zakai} equation as discussed above $-$ with regard to the arbitrary function $\mathscr{f}^{(in)} (\cdot)$ and the Winner random walk process $\mathcal{W}^{(in)}_{i,0}$.
\item {{(\textit{ii}) Outer MFG.}} Defining the state $\mathcal{X}^{(out)}_i(t) \triangleq \mathscr{L}_{\alpha}^{(i)}(t)$ in relation to the $i-$th agent, we have the control law as the game dynamics as follows $d \mathcal{X}^{(out)}_i(t) = \mathscr{f}^{(out)}\big ( \mathcal{X}^{(out)}_i(t)  \big ) dt+ \mathcal{W}^{(out)}_{i,0}$ with regard to the arbitrary function $\mathscr{f}^{(out)} (\cdot)$ and the Winner random walk process $\mathcal{W}^{(out)}_{i,0}$. 
\end{itemize}
\textsc{\textbf{Remark 2.}} \textit{Hereinafter, let us the following holds: $(\cdot)^{(case)} \in \left \{    (\cdot)^{(in)},(\cdot)^{(out)}  \right \}$.}
\textsc{\textbf{Remark 3.}} \textit{The terms $\mathscr{f}^{(in)}\big ( \mathcal{X}^{(in)}_i(t)  \big ) $ and $\mathscr{f}^{(out)}\big ( \mathcal{X}^{(out)}_i(t)  \big ) $ physically consist of the evolutions to respectively $\mathcal{X}^{(in)}_i(t) $ and $ \mathcal{X}^{(out)}_i(t)  $ with regard to the relative gradients. In addition, the amount of the aforementioned divergences and evolutions are upper-bounded, according to the concentration-measure as}
$
\;\mathbb{P}\bigg\lbrace\Big |\mathcal{X}^{(case)}_i(t) {\displaystyle \pm }    \mathscr{f}^{(case)}\big ( \mathcal{X}^{(case)}_i(t)  \big ) \Big|  \le \varkappa _1^{(i,t) }\bigg\rbrace \le \varkappa _2^{(i,t) },
$
\textit{w.r.t. concentration-measure thresholds $\varkappa _1^{(i,t) }$ and $\varkappa _2^{(i,t) }$, correspondingly.}
\textsc{Step 2: Cost.} 
For the trajectory $\left  \{    \mathcal{X}^{(case)}_i(\tau)   \right \} _{[\mathcal{T}^{(case)}_{start},\mathcal{T}^{(case)}_{end}]}$, the cost $\vartheta^{(case)} \big( \mathcal{X}^{(case)}_i;\mathcal{X}^{(case)}_{-i};t  \big)$ consists of the term $\mathscr{g}^{(case)} \big( \mathcal{X}^{(case)}_i(\tau);\mathcal{X}^{(case)}_{-i}(\tau)  \big)$ with given states of other agents, i.e., $\mathcal{X}^{(case)}_{-i}$, and with given initial and final time instants $\mathcal{T}^{(case)}_{start}$ and $\mathcal{T}^{(case)}_{end}$, where we have 
$
\; \vartheta^{(case)} \big( \mathcal{X}^{(case)}_i ;\mathcal{X}^{(case)}_{-i};t\big) \triangleq \mathbb{E} \Bigg(  \int_{\mathcal{T}^{(case)}_{start}}^{\mathcal{T}^{(case)}_{end}}\cdots $$  {\mathscr{g}^{(case)} \big( \mathcal{X}^{(case)}_i(\tau);\mathcal{X}_{-i}^{(case)}\big)
}  d \tau \Bigg)$$=
\mathbb{E} \Bigg(  \int_{\mathcal{T}^{(case)}_{start}}$$^{\mathcal{T}^{(case)}_{end}}\cdots $$ \bigg( c^{(case)}_1 \zeta^{(case)}_{local}$$ \big( \mathcal{X}^{(case)}_i(\tau)\big) $$+$$c^{(case)}_2 \zeta^{(case)}_{global} $$\big( \mathcal{X}^{(case)}_i(\tau)$$;\mathcal{X}^{(case)}_{-i}(\tau) \big)+c^{(case)}_3 \bigg )
d \tau \Bigg) ,
$
where $c^{(case)}_i, i \in \{ 1, 2, 3\}$ are positive constants, and the terms $\zeta^{(case)}_{local}$ and $\zeta^{(case)}_{global}$ stand respectively for the local and global costs.
\textsc{Step 3: HJB \& FPK.} Regarding the fact that we concentrate on the following problem
$
\; \mathop{{\rm min}}\limits_{\ell{\big(\theta^{\star}} \big)} \mathbb{E} \Bigg (   \int_{\mathcal{T}^{(case)}_{start}}^{\mathcal{T}^{(case)}_{end}}  \bigg ( c^{(case)}_1 \zeta^{(case)}_{local} \Big( \ell{\big(\theta^{\star}} \big);\mathcal{X}^{(case)}_i(\tau)\Big) +$$c^{(case)}_2 \zeta^{(case)}_{global} \Big( \ell{\big(\theta^{\star}} \big);\mathcal{X}^{(case)}_i(\tau);\mathcal{X}^{(case)}_{-i}(\tau) \Big)+c^{(case)}_3
 \bigg )d \tau \Bigg ),$$
s.t. $$d \mathcal{X}^{(case)}_i(t) = \mathscr{f}^{(case)}\big ( \mathcal{X}^{(case)}_i(t)  \big ) dt+ \mathcal{W}^{(case)}_{i,0},$$
$
one can integrate the both phrases given above in terms of 
$
\; \mathop{{\rm min}}\limits_{\ell{\big(\theta^{\star}} \big)} c^{(case)}_1 \zeta^{(case)}_{local} \Big( \ell{\big(\theta^{\star}} \big);\mathcal{X}^{(case)}_i(t)\Big) $$+c^{(case)}_2 \zeta_{global} \Big( \ell{\big(\theta^{\star}} \big);\mathcal{X}^{(case)}_i(t);\mathcal{X}^{(case)}_{-i}(t) \Big)+c^{(case)}_3 \;  \\+ \mathbb{E} \;\Big(\vartheta \big( \mathcal{X}^{(case)}_i +d\mathcal{X}^{(case)}_i ;\mathcal{X}^{(case)}_{-i}+d\mathcal{X}^{(case)}_{-i};t+dt\big) \Big),
$
while using the second order Taylor expansion for the expected-value term achieved above $-$ since a non-linear function can be considered a linear one in the context of its projection over a specific horizon.
 $-$ one can mathematically write HJB as
$
\; \partial_t \vartheta^{(case)} +$$ \mathop{{\rm min}}\limits_{\ell{\big(\theta^{\star}} \big)} \Big( \mathcal{X}^{(case)}_i+\ell{\big(\theta^{\star}} \big)\partial_{\mathcal{X}^{(case)}_i} \vartheta^{(case)}  \Big) + \mathcal{W}^{(case)}_{i,1},
$
with the given Winner random walk process $\mathcal{W}^{(case)}_{i,1}$, while FPK is written as
$
\; \partial_t \mathcal{M}^{(case)}=\partial_{\mathcal{X}^{(case)}_i}  \Big( \ell{\big(\theta^{\star}} \big) \mathcal{M}^{(case)} \Big) + \mathcal{W}^{(case)}_{i,2},
$
with the given Winner random walk process $\mathcal{W}^{(case)}_2$, while $\mathcal{M}^{(case)}$ is the probability distribution function (PDF) of the population.
\textsc{Step 4: Convergence rule.} The whole updating process will be automatically terminated when the following condition is satisfied
$
\; \sum\limits_{i} \big|d\mathcal{X}^{(case)}_i  \big| \le \gamma^{(case)},
$
where the term $\gamma^{(case)}$ is a predefined threshold as well, by which the convergence of the relative algorithm is principally guaranteed. 
\textsc{\textbf{Remark 4.}} \textit{In parallel with the inner game, since we have the term $\frac{d \ell \big ( \theta | \mathscr{s}\big )}{d \ell \big ( \theta | \mathscr{s}_0 \big )}$ as discussed in the previous proof, we can conclude that we can even interpret our main problem in terms of a nested 3-level MFG as follows. In fact, the aforementioned term indicates that, in the context of a hidden game, we have some agents as the observations while the initial condition $\ell \big ( \theta | \mathscr{s}_0 \big )$ is the major-player. Therefore, one should take into account a threshold for the maximum divergence between every user with the major-player as $\frac{d \ell \big ( \theta | \mathscr{s}\big )}{d \ell \big ( \theta | \mathscr{s}_0 \big )} \le \eta$. The issue expressed here also implicitly emphatically proves that $\mathcal{M}^{(in)}$ is 2-dimensional. 
Indeed, there exists a 3-level online stochastic optimisation as
$
\; \mathop{{\rm min}}\limits_{\ell\big ( \theta\big)} \mathbb{E}_{\theta} $$\bigg( \Omega_1 \Big(\ell\big ( \theta\big), \mathscr{L}_{\alpha} \Big)  \Big|\mathscr{H}_1\bigg)$$s.t.$$\mathop{{\rm min}}\limits_{\theta}$$ \mathbb{E}_{t} \Big($$\Omega_2 \big( $$ \theta,t  \big)   \Big|\mathscr{H}_1\Big)$$
 \mathop{{\rm min}}\limits_{\ell  \big ( \theta | \mathscr{s}_0 \big ),\ell  \big ( \theta  \big )} $$\mathbb{E}_{\mathscr{S}_0} $$\Big(\Omega_3 $$\big(  \ell  \big ( \theta | \mathscr{s}_0 \big ),$$\ell  \big ( \theta  \big ),t  \big)  $$ \Big|\mathscr{H}_1\Big),
$
w.r.t. the given $\Omega_1(\cdot)$, $\Omega_2 (\cdot)$ and $\Omega_3 (\cdot)$, and conditioned on the occurrence of the true hypothesis $\mathscr{H}_1$.}}$\; \; \; \blacksquare$



\section{Proof of Proposition 3:\\ Stabilizability}
\label{sec:C}
\textsc{Solution 1}.

The information cannot be completely and throughly leaked. This is because of the fact that although the data processing inequality says the information leakage is non-negative, when the system experiences a critical point, of course it can be non-impossibly automatically shut down $-$ according to Corollary 1 and the thrid law of thermodynamics.

Now, one can define the suspension dynamics of the overall system. Indeed, it has been theoretically shown that\footnote{See e.g. \cite{19}.} the function $\mathscr{Sus}(\tau,\mathscr{w})$ mathematically expresses the suspension dynamics w.r.t. the following partial derivate equation 
\begin{equation*}
\begin{split}
\; \xi \frac{\partial}{\partial \tau}  \Bigg(  f(\mathscr{Sus}) +\mathscr{E} \frac{\partial \mathscr{Sus}}{\partial \tau}  \Bigg)=\frac{\partial^2 \mathscr{Sus}}{\partial \mathscr{w}^2} ,
\end{split}
\end{equation*}
w.r.t. the boundary conditions $\frac{\partial \mathscr{Sus}}{\partial \mathscr{w}} (\tau,0)=0$ and $\mathscr{Sus}(\tau,\mathscr{w}_0)=\sigma$ where $\xi$ is the density spectra of the $\alpha-$leakage in the system, $\frac{1}{\mathscr{E}}$ stands for the medium amount of the elasticity\footnote{Resilience: As the degree to which a cost is updated in response to price or income changes.} or in other words, $\mathscr{E}$ stands for the viscosity, and $f(\mathscr{Sus})$ expresses the correlation of the \textit{shear-rate} on the shear stress $-$ such as in the \textit{Couette flow} $-$ where the relevance belongs to the gradient of the estimated version of the uncertainties about Byzantines. 

\textbf{\textsc{Remark 5.}} \textit{Please note that the term $\ell (\cdot)$ actualises the \textit{shear-stress} and the relative reformation of the statistical geometry.}

Now, according to the zeroth law of thermodynamics we continue. This law says that when two isolated sub-system separately experience an equilibrium with a given system, consequently, one can say that the two aforementioned sub-systems are in an equilibrium with each other as well. Indeed, according to the zeroth law of thermodynamics, we see that in the suspension case, the inertia forces discussed in Proposition 1 and Appendix \ref{sec:A}, are in equilibrium with each other where these isolated sub-systems can be interpreted according to the Divergence theorem where the total gradient of each sub-part or sub-surface plays the key role for the total system. 

The proof is now completed $-$ although it can be evaluated with the following solution as well.

\textsc{Solution 2}.

From Definition 1, one can see that $\mathop{{\rm min}}\limits_{\ell{\big(\theta^{\star}} \big)} \mathop{{\rm max}}\limits_{\ell{\big(\theta^{\star}|\mathscr{s}} \big)}  \{\cdots \}$ is a non-increasing operand since $\ell{\big(\theta^{\star}} \big)$ which experiences a minimisation, is a function of $\ell{\big(\theta^{\star}|\mathscr{s}} \big)$. Therefore, the temporal rate of the $\alpha-$leakage is non-positive, specifically regarding the fact that the total rate of the system should be non-negative $-$ as discussed above.

The proof is now completed.$\; \; \; \blacksquare$

\section{Proof of Proposition 4:\\ Nash-equilibrium}
\label{sec:D}
 In consequence of the stabilizability, the issue of Nash-equilibrium is also strongly provable. In other words, even this issue is provable since there exists some specific cases between two groups of the Byzantines and the fusion centre where they experience a zero-sum game the final result of which is: either (\textit{i}) lose-win, either (\textit{ii}) win-lose, or (\textit{iii}) half-half, where the later forms an equilibrium. In addition, the Nash-equilibrium has the probability order of bound $\mathbb{P} \Big \lbrace \big |\nu_2 \le \theta^{\star} \le \nu_3 \big | \le \lambda_1 \Big \rbrace \le \lambda_2$, according to the concentrayio-measure principle and w.r.t. the concentration-measure thresholds $\lambda_1$ and $\lambda_2$ and w.r.t the thresholds $\nu_2$ and $\nu_3$ in relation to the true-and-false hypothesis.

The proof is now completed.$\; \; \; \blacksquare$

\section{Proof of Proposition 5:\\ Curl for the vector field $\mathscr{L}_{\alpha}$}
\label{sec:E}
The $\alpha-$leakage $\mathscr{L}_{\alpha}$ is a vector field theoretically consisting of $\ell\big( \theta  \big)$, $\ell\big( \mathscr{s}  \big)$, $\ell\big( \mathscr{s} _0 \big)$, $\ell\big(\mathscr{s}| \theta  \big)$, $\ell\big( \theta | \mathscr{s}  \big)$, $\ell\big( \theta |\mathscr{s}_0  \big)$, $\ell\big( \mathscr{s}_0 | \theta \big)$, $\ell\big( \mathscr{s} |\mathscr{s}_0  \big)$, $\ell\big( \mathscr{s}_0 | \mathscr{s} \big)$, and also $\ell\big( \theta  | \mathscr{Q}\big)$, $\ell\big( \mathscr{s}  |\mathscr{Q}\big)$, $\ell\big( \mathscr{s} _0|\mathscr{Q} \big)$, $\ell\big(\mathscr{Q}| \theta  \big)$, $\ell\big( \mathscr{Q} | \mathscr{s}  \big)$, $\ell\big( \mathscr{Q} |\mathscr{s}_0  \big)$ w.r.t. the arbitary variable $\mathscr{Q}$.

Now, the Stoke's theorem in relation to the curl principle indiates that for the given surface $\psi  \supseteq   \partial \psi$ and the given closed-loop $\partial \psi  \subseteq   \psi$ over this surface\footnote{That is, the enclosed boundary curve.}, one can write the following for the line $\partial\psi_0 \subseteq \partial\psi  $ and the surface $\psi_0 \subseteq \psi $
\begin{equation*}
\begin{split}
\; \iint\limits_{\psi} \big(\nabla \times \mathscr{L}_{\alpha}) . d \big(\psi_0\big)=\int\limits_{\partial\psi}  \mathscr{L}_{\alpha}. d  \big(\partial\psi_0\big) ,
\end{split}
\end{equation*}
where 
\begin{equation*}
\begin{split}
\; \iint\limits_{\psi \triangleq h \big(  \ell(\theta^{\star}|\mathscr{s}), \ell(\theta|\mathscr{s})  \big)} \big(\nabla \times \mathscr{L}_{\alpha}) . d \Big(\psi_0 \triangleq h_0 \big(  \ell(\theta|\mathscr{s}), \ell(\theta |\mathscr{s}_0)  \big)\Big)=\\ \int\limits_{\partial\psi \triangleq  \ell(\theta|\mathscr{s}_0)  }  \mathscr{L}_{\alpha}. d  \big(\partial\psi_0 \triangleq\ell(\theta)\big) ,\;\;\;\;\;\;\;\;\;\;\;\;\;\;\;\;\;\;\;\;\;\;\;\;\;\;\;\;\;\;\;\;\;\;\;\;\;\;
\end{split}
\end{equation*}
holds, as a case in point, for the surface $h \big(  \ell(\theta^{\star}|\mathscr{s}), \ell(\theta|\mathscr{s})  \big)  \supseteq   \ell(\theta|\mathscr{s}_0) $ and the enclosed closed-loop $\ell(\theta|\mathscr{s}_0) \subseteq   h \big(  \ell(\theta^{\star}|\mathscr{s}), \ell(\theta|\mathscr{s})  \big)$ over this surface, and for the sub-loop $\ell(\theta) \subseteq \ell(\theta|\mathscr{s}_0)  $ and the enclosed sub-surface $ h_0 \big(  \ell(\theta|\mathscr{s}), \ell(\theta |\mathscr{s}_0)  \big) \subseteq h \big(  \ell(\theta^{\star}|\mathscr{s}), \ell(\theta|\mathscr{s})  \big) $ for the given functions $f(\cdot)$, $h(\cdot)$ and $h_0(\cdot)$.

\textbf{\textsc{Remark 6.}} \textit{Regarding the fact that the likelihood functions have the polar structure of $re^{i\phi}$ for the given $r \in [0,1]$ and $\phi \in [0,2 \pi]$ where $i \triangleq \sqrt{-1}$, the divergence of $\mathscr{L}_{\alpha}$ in a given point $\theta_{\beta}$ w.r.t. $\ell\big( \theta  \big)$, $\ell\big( \mathscr{s}  \big)$, $\ell\big( \mathscr{s} _0 \big)$, $\ell\big(\mathscr{s}| \theta  \big)$, $\ell\big( \theta | \mathscr{s}  \big)$, $\ell\big( \theta |\mathscr{s}_0  \big)$, $\ell\big( \mathscr{s}_0 | \theta \big)$, $\ell\big( \mathscr{s} |\mathscr{s}_0  \big)$, $\ell\big( \mathscr{s}_0 | \mathscr{s} \big)$, $\ell\big( \theta  | \mathscr{Q}\big)$, $\ell\big( \mathscr{s}  |\mathscr{Q}\big)$, $\ell\big( \mathscr{s} _0|\mathscr{Q} \big)$, $\ell\big(\mathscr{Q}| \theta  \big)$, $\ell\big( \mathscr{Q} | \mathscr{s}  \big)$, $\ell\big( \mathscr{Q} |\mathscr{s}_0  \big)$  is given as  }
\begin{equation*}
\begin{split}
\; \oiint\limits_{\Gamma} \bigg \lbrace \frac{\partial \mathscr{L}^{(\epsilon)}_{\alpha}}{\partial \epsilon}+\frac{1}{\epsilon}\partial \mathscr{L}^{(\epsilon)}_{\alpha}+\frac{1}{\epsilon}\frac{\partial \mathscr{L}^{(\epsilon)}_{\alpha}}{\partial \phi} \bigg \rbrace d \Gamma_0 ,
\end{split}
\end{equation*}
\textit{where the key terms $\epsilon$ and $\phi$ are respectively the diameter function and the angle function relating to the diameters and angles of all the likelihood functions defined above, physically including the test point $\theta_{\beta}$, while $\Gamma \supseteq \Gamma_0$ is the test surface consisting of every 2 curves being defined from the likelihood functions over which the normal vectors are correspondingly defined. }

 \textbf{\textsc{Remark 7.}} \textit{It should also be careful that the divergence of the density of a fluid is zero iff (if and only if) there exists a source or sink $-$ as more-or-less discussed before in the context of the third law of thermodynamics.}

The proof is now completed.$\; \; \; \blacksquare$

\section{Proof of Proposition 6: \\Ascending/descending nature}
\label{sec:F}
First, we know we have to classify the agents into two groups of Byzantines and good agents since Byzantines aim at maximising the error $\mathbb{P}_{\mathbb{e}}$ or equivalently 
\begin{equation*}
\begin{split}
\; \mathcal{D}_{\mathscr{kl}} \Big \lbrace   \mathscr{H}_0  \big|\big| \mathscr{H}1   \Big \rbrace \triangleq   \ell \Big(   \theta \big | \mathscr{H}_1   \Big)  \frac{log \ell \Big(   \theta \big | \mathscr{H}_1   \Big)}{log \ell \Big(   \theta \big | \mathscr{H}_0   \Big)},
\end{split}
\end{equation*}
whereas good agents follow the reverse, while the term $\mathcal{D}_{\mathscr{kl}} \big \lbrace \cdot \big|\big| \cdot \big \rbrace  $ stands for the information theoretic principle of the \textit{Kullback-Leibler Divergence}.

On the other hand, we see that for every pair $\big(  \ell(\cdot),\ell(\cdot)  \big)$ which mathematically creates a surface, the change in the threshold $\nu_1$ relating to the hypothesis pair $\big(   \mathscr{H}_1 , \mathscr{H}_0 \big)$ results in a change in the size of the relative enclosed surface, and consequently, in the size of the closed-loop over which and according the Stokes' theorem we apply an integral. Therefore, we can see that for each group of Byzantines and good agents separately, the likelihood functions are either ascending or descending which actualise respectively a maximisation or a minimisation.

\textsc{\textbf{Remark 8.}} \textit{In relation to the changes of the relative closed-loops $-$ discussed above $-$ enclosed to the likelihood functions, the following analysis can be applied. Through every given Riemannian manifold, one can define a Wilson-loop\footnote{See e.g. \cite{555}, in order to understand what it is.} for which it is satisfied a differential equation of the Schrodinger type. The aforementioned differential equation can justify the positive/negative evolution of the relative closed-loop.}

The proof is now completed.$\; \; \; \blacksquare$

\section{Proof of Proposition 7:\\ Probabilistic formalisation of the true hypothesis $\mathscr{H}_1$}
\label{sec:G}
The proof is provided in the context of the following multi-step solution. 

\textsc{Step 1.}

As proven e.g. in \cite{20}, the following holds
\begin{equation*}
\begin{split}
\; \mathscr{f} \big( \mu_1 \big) = \mathop{{\rm inf}}\limits_{\Omega_1 \triangleq \big \lbrace \Omega: \mathscr{L}_{\alpha} \setminus \{ 0 \} \big \rbrace, \;\mathbb{N}ull \big \lbrace \Omega \setminus 0 \big \rbrace \neq \emptyset  }\frac{\int \limits_{\Omega} curl \big ( \mathscr{L}_{\alpha} \big) d \Omega}{\mathscr{L}_{\alpha}},
\end{split}
\end{equation*}
where $\mathscr{f} \big( \mu_1 \big) $ is technically a function of the first eigenvalue $\mu_1$. It should also be noticed that the above equation is valid when $\mathscr{L}_{\alpha} \neq 0$, since as declared in the equation, the search region emphasis this issue.

\begin{figure}[t]
\centering
\subfloat{\includegraphics[trim={{24 mm} {59 mm} {216mm} {55mm}},clip,scale=0.46]{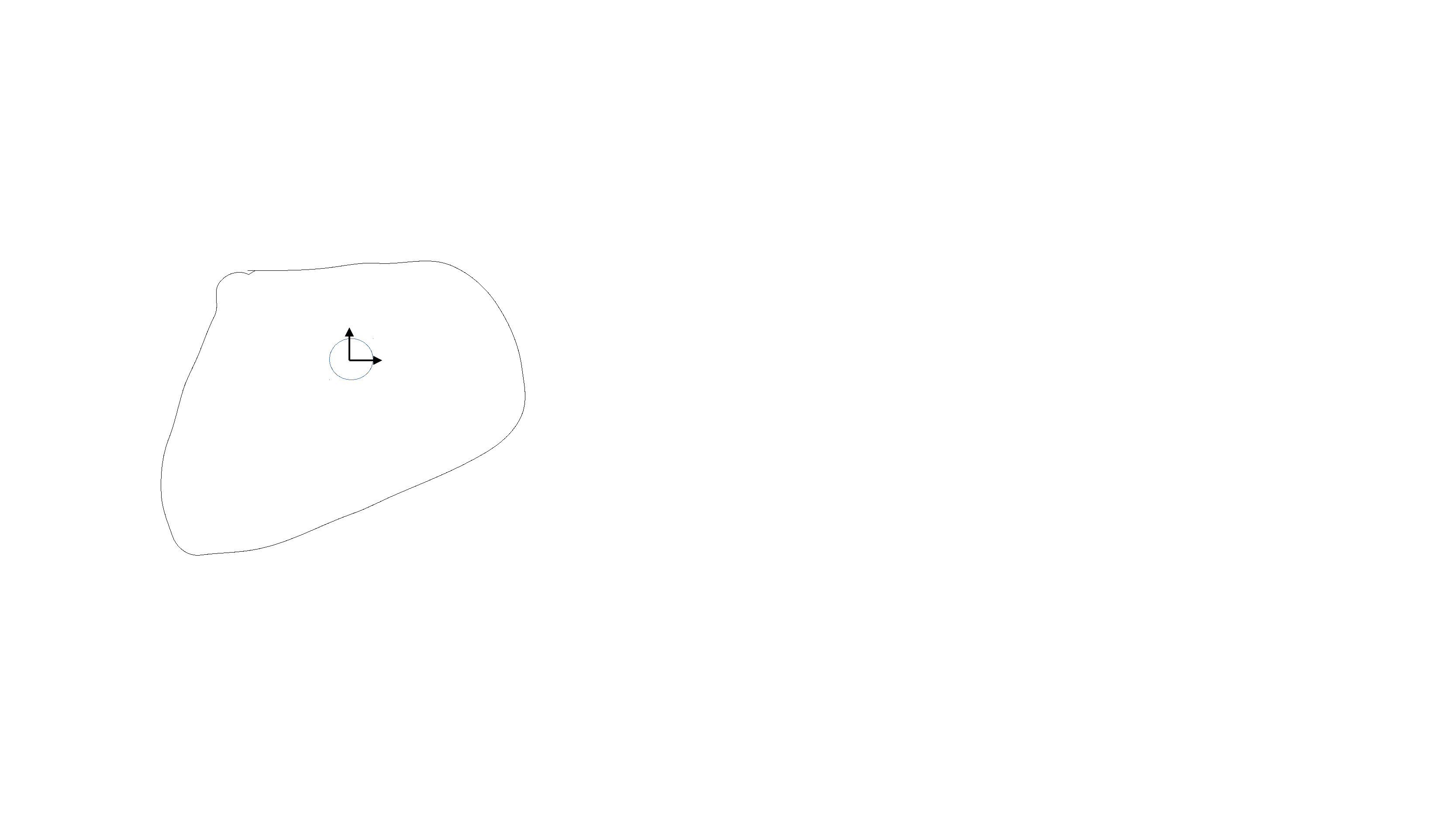}}
\caption{How to hear the orientability of the enclosed surfaces.} \label{ff1}
\label{fig:EcUND} 
\end{figure}

\textsc{\textbf{Remark 9.}} \textit{As perfectly shown in the literature e.g. in \cite{21}, the first-eigenvalue is the central one that plays a totally important role in the creation of a Riemannian manifold. Meanwhile, as proven e.g. in \cite{33}, the facts $\mu_1 \big(  \mathbb{B}\mathscr{all}^{(0)} \big) \le f\Big( \big(     \mu_1(\Upsilon)   \big) \Big)$ and $ \mathbb{B}\mathscr{all}^{(0)}=\mu_1 \big(  \mathbb{B}\mathscr{all}^{(0)} \big) \partial  \mathbb{B}\mathscr{all}^{(0)} ,  \Upsilon=\mu_1(\Upsilon)   \partial \Upsilon $ hold since we theoretically have $\partial  \mathbb{B}\mathscr{all}^{(0)} \le f \Big (  \partial  \Upsilon \frac{\mathbb{V}ol \;\big (  \mathbb{B}\mathscr{all}^{(0)}\big )}{\mathbb{V}ol \; \big ( \Upsilon \big )}   \Big)$ and w.r.t. the sample ball $\mathbb{B}\mathscr{all}^{(0)} \in \mathbb{R}^3$ out of the total manifold $\Upsilon \in \mathbb{R}^3$ where $\partial \Upsilon  \in \mathbb{R}^2$ and $\partial  \mathbb{B}\mathscr{all}^{(0)} \in \mathbb{R}^2$ stand respectively for the surface areas enclosed with the relative volumes $\Upsilon$ and $\mathbb{B}\mathscr{all}^{(0)} $. This kind of interpretation can also be originated from the principle ''hearing the shape of a drum'' invented by M. Kac 1966 \cite{34} where one can infer the shape of the manifold w.r.t. the frequencies of the enclosed susrface area, that is, the eigenvalues of the  Laplacian over the manifold. Finally speaking, it should also be emphatically taken into account that the amount of the orientability named above is upper-bounded. This is because of the fact that the spectral radius, that is, the first eigenvalue theoretically experiences\footnote{See e.g. \cite{gg}} $\mu_1 \le \mathop {max}\limits_{v_1,v_2 } \Big \lbrace \mathscr{d}_{v_2 }+\mathscr{d}_{v_2 }  \Big \rbrace $, with equality if and only if the relative graph relating to the manifold is a regular or semi-regular bipartite one, where the pair of $\big( v_1,v_2 \big)$ stands physically for two vertices with the debsity $\mathscr{d}_{v_1 }$ and $\mathscr{d}_{v_2 }$, respectively. The discussion given here is shown in Fig. \ref{ff1}.}

\textsc{Step 2.} 

Additionally, as discussed in Appendix \ref{sec:F}, the flexibility of the controller-threshold $\nu_1$ may result in a change in the size of the relative surfaces and closed-loops created by the likelihood functions as well. In other words, the resultant vector born from the curl can theoretically change w.r.t. the threshold $\nu_1$. Therefore, one can map the true hypothesis $\mathscr{H}_1: \theta \ge \nu_1$ to a probabilistic one as $\mathscr{H}_1:\mathbb{P} \big \lbrace curl \big ( \mathscr{q}  \big) > \nu_4 \big \rbrace =1-\mathbb{P} \big \lbrace curl \big ( \mathscr{q}  \big) \le \nu_4 \big \rbrace $ where $\mathbb{P} \big \lbrace curl \big ( \mathscr{q}  \big) \le \nu_4 \big \rbrace \triangleq \mathbb{F}_{ curl \big ( \mathscr{q}  \big) }\Big({\nu_4}\Big)$ w.r.t. the CDF $\mathbb{F}_{ curl \big ( \mathscr{q}  \big) } \triangleq \frac{1}{2 \pi i} \oint_{c-i \infty}^{c+i \infty} \frac{-1}{s} x^{-s} \Delta_1 (s+1)ds$ according to the Mellin transform \cite{22} where $\Delta_1(s)=\mathbb{E}_{curl \big ( \mathscr{q}  \big) } \bigg \lbrace \Big (  curl \big ( \mathscr{q}  \big)  \Big )^{s-1} \bigg \rbrace$, while $\mathscr{q}$ may be every possible vector field relating to the surfaces enclosed to the relative likelihood functions. 

\begin{corollary}{\footnote{See e.g. \cite{20, 21, 33, 34}.} For the evolution of the first-eigenvalue, we have the control-law $ \frac{\partial\mu^{\prime}_{1,r}}{\partial t}+\mathscr{A}_0 \mu^{\prime}_{1,r}(t)+\mathscr{A}_1\mu^{\prime}_{1,r}(t-1)=0$ where $\mu^{\prime}_{1,r} \triangleq \mu_{1,r}e^{\mu_1t}$ and $\mu_{1,r}$ is the right-hand eigenvector in relation to the first eigenvalue $\mu_1$.}\end{corollary}

\textsc{\textbf{Remark 10.}} \textit{The corollary defined above also proves a hidden control-law according to which one can extend our 3-level nested game to a 4-level one.}

The proof is now completed.$\; \; \; \blacksquare$

\section{Proof of Proposition 8:\\ Probabilistic solution to the asynchronous behaviour of the Byzantines}
\label{sec:H}
The proof is totally easy to follows which is given in this part, prior to which we go as follows. The pioneer research \cite{qqq} indicates that: ‘’In a synchronous system, processes run in lock step, and messages sent in one step are received in the next.’’ In other words, the term of asynchronous indicates that each
process may be unknown to the other processes since it can go on at its own speed. In fact, one can say that there exists a bounded time delay for message transmission and one can technically guarantee the information as no message is received.

Now, the rate of the $\alpha-$leakage is a perfect tool to probabilistically cope with the asynchronous behaviour of the Byzantines where this rate should be instantaneously negative as much as possible. Indeed, we follow $\mathop{{\rm argmin}}\limits_{\theta}\mathbb{P}r \Big ( \frac{d \mathscr{L}_{\alpha}}{dt} \ge 0 \Big)$, or equivalently, $\mathbb{P}r \Big ( \frac{d \mathscr{L}_{\alpha}}{dt} \ge 0 \Big) \le \gamma_{th}$ w.r.t. the threshold $ \gamma_{th}$, something that can be theoretically actualised by the greedy Algorithm \ref{fo} as well.

\begin{algorithm}
\caption{\textcolor{black}{A Greedy algorithm to cope probabilistically with the asynchronous behaviour of the Byzantines.}}\label{fo}
\begin{algorithmic}
\STATE \textbf{\textsc{Initialisation:}}

(\textit{i}) Invoke $\frac{d\ell  \big ( \theta | \mathscr{s}_0 \big )}{d\ell  \big ( \theta  \big )}$ from $\mathbb{F}$unction $dog-leg$.  ${\;\%\; Algorithm \; \ref{ffff}} \cdots$

(\textit{ii}) Initialise the ground set $\Xi \gets\{ 0,1,2, \cdots, \frac{1}{ \varsigma }-1 \}$, the feasible set $\Theta \gets \emptyset$, and the objective function $f(\Theta) \triangleq \mathbb{P}r \Big ( \frac{d \mathscr{L}_{\alpha}}{dt} \ge 0 \Big)$.

$\;\;$$\;\;$ \textbf{while} $ |f(\Theta) |< f_{max}$ \textbf{do}

$ \; \; \; \; \; \; \; \; \;\;\;\;\;\;\; \Theta_{new} \gets \mathop{{\rm max}}\limits_{{\theta} \; \in \; \Xi\; \setminus \; \Theta} {\rm \; }f( \Theta \; \cup \;
\{ \theta \} )-f( \Theta ), $

$ \; \; \; \; \; \; \; \; \;\;\;\;\;\;\; \Theta \gets \Theta\cup \{ \Theta_{new} \} $

$\;\;$$\;\;$ \textbf{endwhile} 

\textbf{\textsc{Output:}} $\Theta$

\textbf{end}

\end{algorithmic}
\end{algorithm}

\textsc{\textbf{Remark 11.}} \textit{The total mount of the complexity has the order of bound $\mathcal{T} \times \Big(\mathcal{O}\big(\rho^{-1}\big)+\mathcal{O}\big(\rho^{-2}\big)+(NlogN)^2 \Big)$ where $\mathcal{T}$ stands literally for the overall running time.}

The proof is now completed.$\; \; \; \blacksquare$

\begin{algorithm}
\caption{\textcolor{black}{An SPSA method based algorithm to cope with the asynchronous behaviour of the Byzantines.}}\label{foo}
\begin{algorithmic}
\STATE \textbf{\textsc{Initialisation.}} 

Invoke $\frac{d\ell  \big ( \theta | \mathscr{s}_0 \big )}{d\ell  \big ( \theta  \big )}$ from $\mathbb{F}$unction $dog-leg$.  ${\;\%\; Algorithm \; \ref{ffff}} \cdots$

$\;\;$$\;\;$ \textbf{while} $ \mathbb{TRUE}$ \textbf{do}

$\; \; \; \; \; \; \; \; \;$ Compute $  f \Big(  \ell ^{(n)}\big(\theta \big) \Big)    $,

$\; \; \; \; \; \; \; \; \;$ Compute $ \nabla f \Big(  \ell ^{(n)}\big(\theta \big) \Big) $,

$ \; \; \; \; \; \; \; \; \;$ Update $ f \Big(  \ell ^{(n+1)}\big(\theta \big) \Big)  \gets  f \Big(  \ell ^{(n)}\big(\theta \big) \Big)  +\nabla f \Big(  \ell ^{(n)}\big(\theta \big) \Big) $.

$\;\;$$\;\;$ \textbf{endwhile} 

\textbf{\textsc{Output:}} $\ell \big(\theta \big) $

\textbf{end}

\end{algorithmic}
\end{algorithm}

\section{Proof of Proposition 9:\\ Solution to the asynchronous behaviour of the Byzantines}
\label{sec:I}
In addition to the first solution given above, there exists another one which is given in the following.

\begin{figure}[t]
\centering
\subfloat{\includegraphics[trim={{124 mm} {64 mm} {21 mm} {36mm}},clip,scale=0.46]{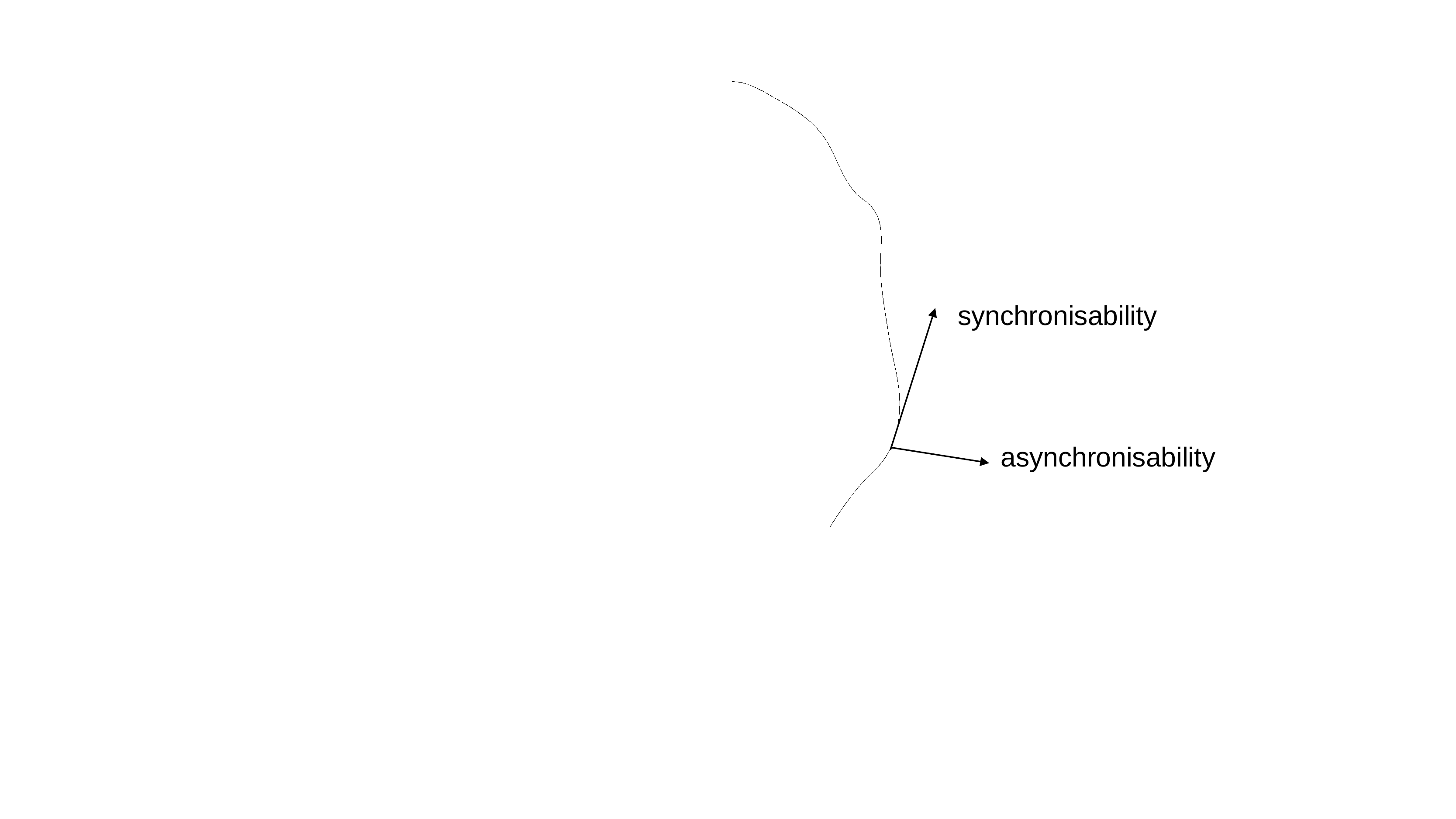}}
\caption{How the asynchronous behaviour of the Byzantines creates an adversarial effect on the algorithm: where the asynchronisability creates a degradation in the gradients.} \label{ff2}
\label{fig:EcUND} 
\end{figure}

As e.g. declared in \cite{q1} and shown in Fig. \ref{ff2}, a synchronous process is the one where $ \mathop {max}\limits_{\mathscr{Q} }    \mathbb{P}r \big(  \mathscr{Q}  \big) \rightarrow 1$ holds w.r.t. the arbitary $\mathscr{Q} $. This means that, after a careful look at the denominator of the equation in relation to the $\alpha-$leakage, it is revealed that a minimisation over the $\alpha-$leakage relies upon a maximisation of the synchronisation of the overall system. Now, the asynchronous behaviour of the Byzantines can result a degradation in the total synchronisability of the system. Therefore, one can assign an unknown random variable $\xi_{b} \in [0,1)$ coordinated by the Byzantines\footnote{As a strictly positive weighting scalar.} where we aim at maximising the synchronisability. This maximisation can be fully performed according to non-convex optimisation algorithms such as the \textit{Simultaneous Perturbation Stochastic Approximation (SPSA)} method\footnote{See e.g. \cite{13, q2} to understand what it is.} as follows.  With the goal of approximately estimating the gradient of the relative objective function, the SPSA method theoretically utilizes finite differences between two randomly perturbed inputs in terms of the following term
\begin{equation*}
\begin{split}
\;\nabla f \Big(  \ell ^{(n)}\big(\theta \big) \Big) \triangleq \frac{f \Big(  \ell ^{(n)}\big(\theta  \big) + \mathscr{C}_{n}\Delta_n\Big)-f \Big(  \ell ^{(n)}\big(\theta  \big) - \mathscr{C}_{n}\Delta_n\Big)}{2\mathscr{C}_{n}\Delta_n},
\end{split}
\end{equation*}
where
\begin{equation*}
\begin{split}
\; f \Big(  \ell \big(\theta^{\star}  \big) \Big) \triangleq \mathop{\rm max}\limits_{\ell \big(\theta^{\star}  \big)}\xi_{b}\mathbb{E}    \left \{ \Big( \ell\big(\theta^{\star}=\theta|\theta \big) \Big )^{\frac{\alpha}{\alpha-1}}\right \}, \xi_{b} \in [0,1),
\end{split}
\end{equation*}
where $\Delta_n$ is a random perturbation vector whose elements are Bernoulli disributed\footnote{See e.g. \cite{13, q2}.}, and w.r.t. the term $\mathscr{C}_{n}$ as the \textit{hyper-parameter} for the SPSA\footnote{See e.g. \cite{13, q2}.}, and the term $n$ is the index of the iteration.

Finally, the algorithm updates as $ f \Big(  \ell ^{(n+1)}\big(\theta \big) \Big)  \gets  f \Big(  \ell ^{(n)}\big(\theta \big) \Big)  +\nabla f \Big(  \ell ^{(n)}\big(\theta \big) \Big) $.

The discussion given theoretically above is presented in Algorithm \ref{foo}.

\textsc{\textbf{Remark 12.}} \textit{The total mount of the complexity has the order of bound $\mathcal{T} \times \Big(\mathcal{O}\big(\rho^{-1}\big)+\mathcal{O}\big(\rho^{-2}\big)+0.33N \Big)$ where $\mathcal{T}$ stands literally for the overall running time. The term $0.33N$ also arises from SPSA\footnote{See e.g. \cite{13, q2}.}.}

The proof is now completed.$\; \; \; \blacksquare$

\section{Proof of Proposition 10:\\ AoI}
\label{sec:J}
In e.g. \cite{qqq2}\footnote{Page $3$, Eq. $3$.}, it was proven that the probability of transmission at the time instant $t+1$ at the $i$ node is given by $\mathscr{P}_{i}(t+1) =max \bigg \lbrace \mathscr{P}^{(min)}_{i}   ,    \mathscr{f} \Big (   \mathscr{P}_{i}(t);   -\frac{1}{\vartheta^{(aoi)}_{i}(t)}  \Big)   \bigg \rbrace$, where $\mathscr{P}^{(min)}_{i}$ is a constant independently and individually pre-adjusted at each node $i$ and $\vartheta^{(aoi)}$ is AoI which we are interested in making a reduction in. 

After a careful look at the equation $max \bigg \lbrace \mathscr{P}^{(min)}_{i}   ,    \mathscr{f} \Big (   \mathscr{P}_{i}(t);  -\frac{1}{\vartheta^{(aoi)}_{i}(t)}\Big)   \bigg \rbrace$, we see that this can be re-presented as $\mathscr{P}^{(min)}_{i} u(t) + \big (t-\mathscr{P}^{(min)}_{i}\big) u \big(t-\mathscr{P}^{(min)}_{i}\big)$ where $u(\cdot)$ is the step function, i.e., $u(t)=1, t \ge 0$ and $u(t)=0, t < 0$ hold. Now, we see that for $\mathscr{P}^{(min)}_{i}  \le    \mathscr{f} \Big (   \mathscr{P}_{i}(t);  -\frac{1}{\vartheta^{(aoi)}_{i}(t)} \Big) $ one can see a direct way to reach out AoI $\vartheta^{(aoi)}$ according to $\mathscr{P}_{i}(t+1) $ and $\mathscr{P}_{i}(t) $ $-$ that is, the two terms that can be found by the temporal rate of the $\alpha-$leakage according to $\mathbb{F}=ma$.

On the other hand, a maximisation over $-\frac{1}{\vartheta^{(aoi)}_{i}(t)}$ means that we should take into account the worst-case of it, i.e., the the value we look for AoI should be the least upper-bound of it $-$ or vice versa.

The proof is now completed.$\; \; \; \blacksquare$

%
\markboth{IEEE, VOL. XX, NO. XX, X 2021}%
{Shell \MakeLowercase{\textit{et al.}}: Bare Demo of IEEEtran.cls for Computer Society Journals}
\end{document}